\documentclass[12pt]{article}

\usepackage[a4paper, margin=0.8in]{geometry}

\usepackage{graphicx} % Required for inserting images
\usepackage{booktabs}
\usepackage{url} 
\usepackage[colorlinks=true,linkcolor=black,citecolor=black,urlcolor=blue]{hyperref}

\usepackage{natbib}

\usepackage{amsmath}
\usepackage{tabularx}

\usepackage{setspace}
\usepackage{bm}
\usepackage{scalerel}
\usepackage{float}

\newcommand{\bbeta}{\bm \beta}
\newcommand{\bmx}{\bm x}

\usepackage{amssymb}  % for checkmark
\usepackage{pifont}  % for cross symbol
\newcommand{\cmark}{\textcolor{green!60!black}{\ding{51}}} % green check mark
\newcommand{\xmark}{\textcolor{red}{\ding{55}}}            % red x mark
\newcommand{\lmark}{\textcolor{yellow!70!black}{\boldmath\(\triangle\)}}
\newcommand{\code}[1]{\texttt{#1}}

\usepackage{rotating}
\usepackage{makecell}
\usepackage[table]{xcolor}
\definecolor{lightgray}{gray}{0.85}

\usepackage{microtype}
\usepackage{tikz}
\usetikzlibrary{shapes.geometric, arrows}
\usetikzlibrary{math} %needed tikz library
\usetikzlibrary{positioning}

\usepackage{pdfpages}

\title{sae4health: An R Shiny Application for Small Area Estimation in Low- and Middle-Income Countries}
\author{by Yunhan Wu$^1$, Qianyu Dong$^2$, Jieyi Xu$^1$, Zehang Richard Li$^2$ and Jon Wakefield$^{1,3}$\\
\vspace{0.2em}\\
$^1$Department of Biostatistics, University of Washington, Seattle, USA\\ 
$^2$Department of Statistics, University of California, Santa Cruz, USA\phantom{te}\\
$^3$Department of Statistics, University of Washington, Seattle, USA \phantom{te}}

\begin{document}

\maketitle

\abstract{
Accurate subnational estimation of health indicators is critical for public health planning, particularly in low- and middle-income countries (LMICs), where data and analytic tools are often limited. sae4health is an open-access Shiny application (\url{https://rsc.stat.washington.edu/sae4health/}) that generates small area estimates for more than 150 demographic and health indicators, based on over 150 Demographic and Health Surveys (DHS) from 60 countries. The platform offers both area- and unit-level models with spatial random effects, implemented through fast Bayesian inference using Integrated Nested Laplace Approximation (INLA). The app is fully browser-based and requires no data input, programming skills, or statistical modeling expertise, making advanced methods accessible to a wide range of users. Estimates are processed in real time and presented as interactive maps, tables, and downloadable reports. A companion website (\url{https://sae4health.stat.uw.edu}) provides documentation and methodological background to support the app. Together, these resources enhance access to subnational health data and facilitate the use of DHS surveys for evidence-based decision making.

}

\section{Introduction}

\label{sec:intro}

Accurate subnational estimates of health and demographic indicators are essential for public health planning in low- and middle-income countries (LMICs), where data gaps often hinder efforts to address regional disparities \citep{wickremasinghe:etal:16}. These estimates are vital for 
cataloging inequalities, allocating resources, assessing interventions, and gauging progress toward targets, such as the United Nations (UN) Sustainable Development Goals (SDGs) \citep{SDG4}. However, obtaining reliable subnational estimates is challenging due to data limitations, complex survey designs, and the need for specialized statistical expertise \citep{tzavidis:etal:18}.

Household surveys, particularly the Demographic and Health Surveys (DHS), provide the most reliable and timely population health data in LMICs. With over $300$ surveys conducted across more than $90$ countries since 1984, DHS datasets offer extensive insights into demographic and health indicators \citep{croft2018guide} and have been widely used to shape policies. According to the database maintained by the DHS program, between October 2003 and April 2025, there have been $6165$ peer-reviewed journal articles based entirely or primarily on DHS data. These papers span a wide range of topics, including  maternal health, childhood mortality, vaccination coverage, fertility, nutrition, malaria, tuberculosis, HIV/AIDS, and education, among others. The DHS surveys have been the main source of data on the general population in most LMICs, providing a uniquely comparable data source covering countries in several continents for over four decades.

Data from DHS surveys are usually distributed as multiple recode data files, with each row corresponding to one case, e.g., household, individual, mother, birth, etc., and each column a variable. From the case-level data, DHS routinely computes over $2500$ health, nutrition, and demographic indicators \citep{croft2018guide}. 
Prevalence estimates of these indicators at the national level (and sometimes at one administrative level below national) are made public freely from the DHS STATcompiler \citep{statcompiler}. These estimates of prevalence are produced following standard design-based approaches, i.e.,survey-weighted estimators. The weighted estimates are also reported for key indicators in DHS survey reports, sometimes by key demographic subgroups.  
DHS currently does not release the case-level processed indicators. \citet{croft2018guide} describes the definition of each indicator and the changes over time. Processing and extracting these indicators from the DHS recode files can be a tedious and error-prone task.

In addition, DHS currently does not produce prevalence estimates at finer geographic resolution. 
Data sparsity at finer resolutions necessitates small area estimation (SAE) techniques. The complex design of DHS, typically a stratified two-stage cluster sample with unequal sampling probabilities, further complicates SAE applications, requiring specialized methods for accurate estimation and valid inference \citep{wakefield2020small,wakefield2020two}. Traditional SAE methods, originally developed for high-income countries, often rely on auxiliary data such as administrative records \citep{rao:molina:15}, but these are frequently unavailable in LMICs, limiting their applicability. Despite these challenges, recent methodological advancements have successfully adapted SAE for key health and demographic indicators in LMICs, including child mortality \citep{golding2017mapping, mercer:etal:15, li:etal:19, wakefield2019estimating, wu:etal:21}, HIV prevalence \citep{wakefield2020small}, and vaccination coverage \citep{utazi2020geospatial,lbd2021measles,dong2021modeling}. These developments highlight the potential for SAE as a scalable and practical tool for public health applications in LMICs.

However, the implementation of these methods requires advanced statistical knowledge and programming skills, limiting their accessibility for analysts and public health officials in LMICs who need timely insights. Existing methodological framework often lack user-friendly implementation, clear practical guidance, and scalability \citep{tzavidis:etal:18,giorgi2021model}. While a few software tools support SAE for household surveys, in particular DHS surveys, they often fall short in one or more critical areas. To be practically useful in LMIC settings, we believe a tool must bring together several key features: 
\begin{enumerate}
  \item being able to produce subnational estimates beyond the admin-1 level,
    \item providing customizable workflows with reproducible results,
    \item include uncertainty quantification, 
    \item allow informative interactive visualization for summarizing and communicating results, 
    \item provide a user interface that straightforward to use for non-statisticians.
\end{enumerate}

Yet, no existing tool fully meets these criteria. To this end, we developed \texttt{sae4health}, a user-friendly R Shiny built on the R package \texttt{surveyPrev}. Designed for public health analysts, policymakers, and researchers with limited programming experience and statistical background, \texttt{sae4health} streamlines subnational estimation and prevalence mapping for over 150 binary demographic and health indicators derived from DHS data. It integrates established statistical methodologies while maintaining accessibility through guided model selection, model fitting, and interactive visualization. Our goal with \texttt{sae4health} package is to 
% bridge the gap between statistical expertise and local knowledge, \texttt{sae4health} 
empower public health officials and data analysts to produce accurate, contextually relevant prevalence estimates of key demographic and health indicators through a guided workflow. 
% This hands-on approach to analysis supports more informed health planning, 
% enabling users to interact with data directly and understand the implications of their results for targeted, effective interventions.

The remainder of this paper is organized as follows. We first introduce the statistical approaches from SAE and computational methods that underpin \texttt{sae4health}. Next, we describe the database for the app and the app’s workflow, detailing its analysis plan for model selection and fitting. We then present a case study on under-five stunting based on 2018 Nigeria DHS, showcasing the \texttt{sae4health}’s key features, including its full suite of interactive visualizations for interpretation, diagnosis, and inference. Finally, we conclude with a discussion of the \texttt{sae4health}’s practical value and its potential extensions.

\texttt{sae4health} is available both as an online application at \url{https://rsc.stat.washington.edu/sae4health} and as a CRAN package for local installation, with full documentation accessible at \url{https://sae4health.stat.uw.edu}.

\section{Statistical Models and Computation}

\subsection{Overview}

Most DHS surveys use a stratified two-stage cluster sampling design with unequal selection probabilities. While a small number of countries deviate from this design, the typical structure involves stratification by a cross-classification of urban/rural areas and the first administrative level below national (Admin-1 regions). In the first stage of cluster sampling, clusters are sampled within strata, using probability proportional to size (PPS) sampling, with the size variable being the number of households. At the second stage, a fixed number of households is probabilistically sampled from those available. All members of the selected household are interviewed.

For illustration, the 2018 Nigeria DHS used the 2006 census as its sampling frame, updated with revised urban/rural classifications and population growth projections. The survey was stratified by urban/rural residence within each of the $37$ administrative units. In the first stage, $1,400$ clusters ($580$ urban, $820$ rural) were selected using probability proportional to size. In the second stage, 30 households per cluster were randomly selected, yielding a total sample of approximately $42,000$ households.

Most of the DHS indicators are binary, which is the main focus of \texttt{sae4health}. The prevalence of an indicator is the proportion of individuals or households that exhibit that outcome among a pre-specified population, within a given area and time period. 

For all indicators, the app supports analysis using three commonly used small area estimation (SAE) methods: direct estimates, area-level, and unit-level models, each suited to different data sparsity and precision needs. Before detailing each method, we summarize their characteristics in Table \ref{tab:sae_comparison}, which highlights their respective strengths and limitations.

\begin{table}[!ht]
\centering
\begin{tabular}{|c|c|c|c|}
\hline
\textbf{Feature} & \textbf{Direct} & \textbf{Area-Level} & \textbf{Unit-Level} \\ \hline
Handling sparse data& \xmark & \lmark & \cmark \\ \hline
Quantifying uncertainty & \cmark & \cmark & \cmark \\ \hline
Spatial smoothing & \xmark & \cmark & \cmark \\ \hline
Design consistency & \cmark &  \cmark & \xmark \\ \hline
Precision of estimates &  \xmark & \lmark & \cmark \\ \hline
Computation complexity &  \cmark & \lmark & \lmark \\ \hline
\end{tabular}
\caption{Comparison of SAE methods in the Shiny app. Symbols: \cmark\ indicates strong suitability, \lmark\ moderate suitability, and \xmark\ low suitability for each aspect evaluated.}
\label{tab:sae_comparison}
\end{table}

The following sections provide a detailed overview of each method, their formulation, and the contexts in which they are most effective. Throughout the discussion of statistical methodology, we assume the DHS recode data can readily be processed into the binary indicator of interest. This step is carried out using the \texttt{surveyPrev} package \citep{surveyPrev}.

\subsection{Direct Estimates}

{Direct estimates} serve as the starting point for analysis, providing a straightforward and design-based approach to inference. The term ``direct'' refers to the fact that estimates for a given area rely solely on the response data from that area, without borrowing information from other regions. This method is particularly useful when the sample size within an area is sufficiently large, as it avoids modeling assumptions and provides inference grounded purely in the survey design.

Let $S_i$ denote the set of sampled clusters within area $i$. Focusing on prevalences, 
% with the target being the region $i$ fraction $p_i=T_i/N_i$, where $T_i$ is the total number of individuals with the characteristic of interest and $N_i$ is the total population.
a common direct estimator is the design-based weighted estimator \citep{hajek1971discussion},
\[
\widehat{p}_i^{\,\text{W}} = \frac{\sum_{c \in S_i} w_{ic} y_{ic}}{\sum_{c \in S_i} w_{ic}},
\]
where \( w_{ic} \) is the design weight (inverse sampling probability) for cluster \( c \) in area \( i \), \( y_{ic} \) is the total number of cases in cluster $c$. Design-based variance estimator for $\widehat{p}_i^{\,\text{W}}$ can be computed using the \texttt{survey} package \citep{surveypackage}.
%ensures that survey weights are properly incorporated, adjusting for unequal selection probabilities that arise due to stratified unequal probability cluster sampling.

One of the main advantages of direct estimation is that it fully accounts for the survey design through weighting, which helps avoid bias from informative sampling. For instance, in many LMICs, urban clusters are oversampled, and design-based estimators correctly account for that through weighting. Additionally, when sample sizes within an area are sufficiently large, direct estimates tend to have desirable frequentist properties, with confidence intervals that are well-calibrated under the asymptotic normality assumption. Because no explicit model is imposed on the data, inference is based on minimal assumptions, making this approach intuitively appealing and straightforward to describe.

Despite these advantages, direct estimation has notable limitations, particularly in areas with small sample sizes. When data are sparse within an area, the associated uncertainty (from which confidence intervals are constructed) can be exceedingly large, rendering the estimates impractical for decision-making. In extreme cases, if there are only a few observations in an area, the point estimate for prevalence may take values of 0 or 1, and  the usual formulas for calculating the standard errors will be unusable, leading to standard error estimates that are either erroneously small or mathematically undefined.  Furthermore, the coverage of the uncertainty intervals relies on  asymptotic normality of the sampling distribution of the estimator. While this assumption holds when sample sizes are large, it may be violated in cases of sparse data.

In practice, direct estimation is generally reliable at the national or Admin-1 level, where sample sizes are typically sufficient for stable estimates, except perhaps in cases where the indicator is extremely rare. However, for finer geographic units, such as Admin-2, the uncertainty associated with direct estimates can be unreliable, and in extreme cases, there may be no clusters in an area. This motivates the need for model-based approaches, which improve precision by leveraging information from neighboring regions.

\subsection{Area-Level (Fay-Herriot) Model}

{Area-level (Fay-Herriot) modeling} \citep{fay:herriot:79} is the most popular SAE approach, particularly when high-quality auxiliary variables are available, due to its ability to improve the precision of estimates while maintaining design consistency. Under this approach, the direct estimates are linked together via a hierarchical model, allowing information to be shared across regions. The underlying rationale is that the similarity of prevalence across the study region may be leveraged to fine-tune the estimates in each area. Since data from all areas are used for estimation in each area, this is an example of the production of an ``indirect'' estimate.

We transform the weighted estimates of prevalence to the entire real line using a logit function, 
% with transformed estimates, denoted 
% \[
% Z_i = \text{logit}\left(\widehat{p}_i^{\,\text{W}}\right). 
% \]
and define the sampling model (likelihood) as the asymptotic normal distribution of the estimator:
\[
\text{logit}\left(\widehat{p}_i^{\,\text{W}}\right) \mid \theta_i \sim N(\theta_i, V_i),
\]
where $V_i$ represents the estimated variance of the logit of the direct estimator. 
The underlying logit-transformed area-level prevalence, \( \theta_i \), is modeled hierarchically as:
\[
\theta_i = \alpha + \bmx_i\bbeta + b_i,
\]
where \( \alpha \) is a global intercept, $\bmx_i$ are area-level covariates, with associated regression coefficients $\bbeta$, and \( b_i \) captures spatial structure through the BYM2 model \citep{besag1991bayesian}, a reparameterization of the Besag-York-Mollié (BYM) model \citep{riebler2016an}. This model decomposes \( b_i \) into an independent normal component \( e_i \) and a spatially structured component \( S_i \):
\[
b_i = \sigma (\sqrt{1-\phi} e_i + \sqrt{\phi} s_i),
\]
where \( s_i \) follows a scaled intrinsic conditional autoregressive (ICAR) prior. Penalized complexity (PC) priors \citep{simpson2017penalising} are applied to the hyperparameters \( \sigma \) (total standard deviation) and \( \phi \) (the proportion of spatial variation). This formulation allows for both local randomness and spatial dependence, ensuring flexibility in capturing area-level variation.

The Fay-Herriot model enhances precision while preserving design consistency by modeling design-based estimates and their variances, avoiding potential bias from informative sampling. By pooling data across areas, it reduces uncertainty, especially in sparsely sampled regions. A key feature is its shrinkage effect, which pulls extreme estimates toward the overall mean or local spatial patterns. While this introduces shrinkage bias, the gain in precision will offset such bias. The computation of Fay-Herriot model is more intensive than direct estimation, but the implementation used in the app is very efficient, and computation time is not prohibitive.

On the disadvantages, Fay-Herriot model depends on the availability and stability of direct estimates and their variances. When data are extremely sparse, direct estimates and the associated uncertainty measures may be unreliable or missing, limiting the model’s applicability. Moreover, interval estimates based on hierarchical models do not typically have the usual frequentist coverage, when each area is viewed in isolation. Across the map the coverage will be close to nominal \citep{burris:hoff:20}, but explaining the uncertainty in each area is more tricky.

At finer geographic levels, such as Admin-2, design-based variance estimates can be unstable and cause numerical issues. A common workaround is excluding areas with unreliable variance estimates, which may introduce slight bias. Despite these challenges, estimates from area-level models consistently show shrinkage relative to direct estimates. However, the shrinkage is quite predictable from known theory, since the Fay-Herriot model is an example of a linear mixed effects model, for which rich theory exists.
%, though less so than unit-level models, which apply finer-scale smoothing.

\subsection{Unit-Level Model}

{Unit-level models}, introduced by \cite{battese:etal:88}, adopt a fully model-based approach by specifying a sampling model at the cluster level, making them distinct from direct and area-level methods, which rely on weighted estimates. These models use an overdispersed binomial distribution to account for within-cluster variability and are particularly valuable when data are sparse. Like the area-level model, the unit-level model incorporates a hierarchical structure that links prevalence estimates across areas.
For a given cluster $c$ within area $i$ prevalence \( p_{ic} \) follows a Beta-Binomial distribution: 
\[
Y_{ic} | p_{ic},d  \sim \text{BetaBinomial}(n_{ic}, p_{ic}, d),
\]
where \( d>0 \) represents the overdispersion parameter. The logit of \( p_{ic} \) includes area-level random effects and potentially area-level auxiliary variables, modeled as \( \alpha + \bmx_i\bbeta+ b_i \), allowing for spatial smoothing across clusters, with $b_i$ again following a BYM2 model. The area-level prevalence is shared across all clusters in the area, such that we have $p_i = \text{expit}(\alpha + \bmx_i\bbeta+b_i)$.

Cluster-level models effectively handle very sparse data and they provide the only method that can be used under such cases, as they do not depend on the existence of direct estimates and their variances, nor the asymptotic properties of direct estimates. When the assumed likelihood model reasonably approximates the true data-generating process, unit-level models make efficient use of the available data. Similar to the area-level models, unit-level models introduce shrinkage bias but compensate by significantly reducing uncertainty. As a result, they achieve lower mean squared error on average, balancing bias and variance to improve overall estimation accuracy.

The use of a sampling model for cluster-level responses fundamentally differentiates this approach from direct estimates and area-level models, which rely on weighted estimates. Validating this assumption is challenging, particularly for rare outcomes. In the absence of survey weights, the unit-level model is subject to systematic bias if the survey design is not adequately accounted for  \citep{wu2024modelling}. In cases of sparse data and rare outcomes, excessive shrinkage may occur, as the limited information may not be sufficient to distinguish local variations, leading to an overly smoothed prevalence surface. Although computationally more intensive than direct estimation, modern Bayesian techniques ensure feasibility for unit-level models, even for large-scale survey data.

\subsection{Computation}

The statistical analysis in \texttt{sae4health} is fully automated to deliver robust Bayesian hierarchical modeling capabilities, while allowing users to perform sophisticated Bayesian modeling with only a few clicks. Behind the scenes, the app relies on the Integrated Nested Laplace Approximation (INLA) method, implemented via the R \texttt{INLA} package \citep{rue2009approximate}, to achieve efficient and accurate posterior inference. Extensive simulations in \cite{osgood2023statistical} illustrate the accuracy of INLA for spatial modeling of the type used in the app. \texttt{sae4health} manages detailed specifications for priors and hyperpriors, appropriately handling both spatially structured and unstructured random effects within spatial smoothing models to ensure reliable model performance.

To support different computational environments, we offer two deployment options. The primary venue is a web-based version hosted on a remote server (available at \url{https://rsc.stat.washington.edu/sae4health}), where all computations are performed on high-speed cores. This option enables users to run the full analysis pipeline entirely through a browser, with no need for local setup, data upload or computational resources, only requiring a stable internet connection. The second option is an R package available on CRAN, which requires local installation and executes all computations and data preparation directly on the user’s machine, which offers greater flexibility and control, but requires user to upload raw DHS data.

\tikzstyle{decision} = [diamond, draw, fill=blue!20, 
    text width=6.5em, text badly centered, node distance=3cm, inner sep=0pt]
\tikzstyle{block} = [rectangle, draw, fill=blue!20, 
    text width=11em, text centered, rounded corners, minimum height=4em]
    \tikzstyle{block2} = [rectangle, draw, fill=green!20, 
    text width=15em, text centered, rounded corners, minimum height=5em]
\tikzstyle{line} = [draw, -latex']
\tikzstyle{cloud} = [draw, ellipse,fill=red!20, node distance=3cm, text width=7.5em,text centered,
    minimum height=2em]
  %  \centering
 %   \begin{small}
\begin{figure} [ht]
    \begin{center}
\begin{tikzpicture}[node distance = 4cm, auto]
    % Place nodes
    \node [block, xshift=2.5cm] (level1) {Sufficient Sample Size at Desired Admin Level?};
  %  \node [cloud, left of=init] (expert) {expert};
        \node [block, below of=level1, yshift=0.5cm, xshift=3.5cm] (level2right) {Reliable Weighted Point and Variance Estimate?};
        \node [cloud, below of=level1, yshift=-0.5cm, xshift=-2cm] (level2left) {Weighted Estimation };
            \node [cloud, below of=level2right, yshift=-0.5cm, xshift=3.5cm] (level3right) {Unit-Level \\ Spatial Model};
        \node [cloud, below of=level2right, yshift=-0.5cm, xshift=-2.5cm] (level3left) {Area-Level \\Spatial Model};

    \path [line] (level1) -- (level2left);
        \path [line] (level1) -- (level2left);
            \path [line] (level2right) -- (level3left);
        \path [line] (level2right) -- (level3right);
  %  \path [line] (identify) -- (evaluate);
 %   \path [line] (evaluate) -- (decide);
    %\path [line] (decide) -| node [near start] {yes} (update);
    %\path [line] (update) |- (identify);
    \path [line] (level1) -- node {no}(level2right);
        \path [line] (level1) -- node {yes}(level2left);
            \path [line] (level2right) -- node {no}(level3right);
        \path [line] (level2right) -- node {yes}(level3left);
 %   \path [line,dashed] (expert) -- (init);
  %  \path [line] (init) -- (system);
  %  \path [line,dashed] (system) |- (evaluate);
\end{tikzpicture}
\caption{Schematic of overview for model selection.}\label{fig:app_decision-making-process}
\end{center}
\end{figure}

\section{Database and Workflow}
 
\subsection{Database}
We host a database on the server used for the primary deployment of the application (online server with link \url{https://rsc.stat.washington.edu/sae4health}). This database stores the essential resources needed to generate DHS-based health indicators on the fly: (1) raw DHS recode data files, (2) harmonized geospatial boundary shapefiles, and (3) indicator coding schemes derived from standardized DHS processing code. Together, these components allow \texttt{sae4health} to dynamically produce analysis datasets for subnational estimation across more than 150 demographic and health indicators from over 150 DHS surveys in 60 countries. We include all DHS surveys conducted after 2000 that have GPS data available, which is crucial for subnational analysis.  

The raw DHS data are distributed in standardized recode formats, each corresponding to a unit of analysis such as households, women, men, births, or children. To compute an indicator, the appropriate recode file(s) must be identified and the correct coding scheme applied. Many indicators can be generated from a single recode, but some require multiple files. For example, overall population HIV prevalence requires combining the HIV recode with both the men’s and women’s recode.  

To systematize this process, we adopt the framework of the \texttt{surveyPrev} package. Coding scheme for each indicator is prepared as individual functions, which are stored internally within the platform. Given a survey and indicator, the corresponding function taking raw DHS data as input can be executed in the backend to produce a clean, analysis dataset. These coding schemes have been validated on several benchmark surveys by comparing app-generated national estimates with the official DHS estimates. In addition, the workflow provides such comparisons on the fly for quality assurance as well. This design ensures reproducibility and shields users from the complexity of manual recode processing.  

Subnational estimation requires data to be geo-indexed at the required level. For this, we rely on administrative boundary shapefiles from the Global Administrative Areas (GADM) database, which provide consistent linkages across administrative levels. DHS cluster-level GPS data are then used to assign survey records to the correct administrative units by matching clusters to their coordinates. This integration is carried out seamlessly within the application workflow.  

In summary, the \texttt{sae4health} database consists of raw DHS recodes, indicator-processing functions, and boundary shapefiles. These components support a fully automated pipeline in which analysis datasets are generated in real time, enabling users to obtain subnational estimates without handling raw data or specialized statistical code.  

\subsection{An Overview of the Workflow}

The interface of \texttt{sae4health} follows a structured workflow to guide users through the exercise of prevalence mapping in the following steps:

\begin{enumerate}
    \item User specifies a survey and
    % after users upload their data and specify 
    one or multiple target inference levels from national (Admin-0) to Admin-2. In some countries, Admin-3 analysis is also possible. 

    \item Data processed to prepare analysis data set. For the web version of the app hosted at \url{https://rsc.stat.washington.edu/sae4health}, all DHS surveys have been stored locally on the server and no data upload is needed. For local version, user uploads the required DHS data recode files following the provided instructions. 
    
    \item Automatic data sparsity assessment are performed for the user-selected model and levels of inference. Initial model recommendations are made based on data availability. User select which models to run.
    \item A series of visualization tools is provided to assess and compare different summaries of estimated prevalences. Users compare fitted models across different levels of inference and may iteratively refine the model choices and specifications.

    \item Users generate a summary report of the selected models. Users can also download static visualizations and prevalence estimates in the process.

\end{enumerate}

% \subsection{Model choice}

A key feature for \texttt{sae4health} is the structured decision-making framework for model selection. 
Figure \ref{fig:app_decision-making-process} illustrates the initial model recommendations based on data sparsity and reliability of the weighted estimates, i.e., step 3 of the workflow. 
% This process is driven by data sparsity and the reliability of survey-weighted estimates and their associated uncertainties. 
While survey and indicator contexts may vary, direct (weighted) estimation is generally feasible at the Admin-1 level due to a sufficient number of observations. However, at the Admin-2 level, alternative approaches are often necessary. When reliable weighted point and variance estimates are available, an area-level spatial model can be applied; otherwise, a unit-level spatial model is required to enhance accuracy.

To assist users, built-in data sparsity warnings identify dataset limitations and provide guidance in specific contexts. \texttt{sae4health} currently employs the following messages:

\begin{itemize}
    \item \textbf{Direct Estimates:} A warning is raised if
    \begin{equation*}
        \texttt{No data areas} + \texttt{Low information areas} > 25\%.
    \end{equation*}
    Here, \texttt{Low information areas} refer to areas with very sparse data or a point estimate of 0 or 1, where standard survey-weighted formula is not usable. In this case, \texttt{No data areas} return \texttt{NA} for both point and uncertainty estimates, while \texttt{Low information areas} report point estimates of 0 or 1 (weighted estimate) but \texttt{NA} for uncertainty. Users can override the warning to obtain direct estimates where data exist. 
    
    \item \textbf{Area-Level Model:} An error is triggered if
    \begin{equation*}
        \texttt{No data areas} + \texttt{Low information areas} > 25\%.
    \end{equation*}
    In such cases, area-level models are not allowed, as sparse data across many areas compromises estimation. If such areas make up $\le 25\%$ of all areas,  data from these areas are excluded before model fitting. 
    % as a pragmatic solution, with 
    Models will then rely on random effects (and covariates if used) to fill in the missing areas. 
    
    \item \textbf{Unit-Level Model:} A warning is issued if:
    \begin{equation*}
        \texttt{No data areas} > 25\%,
    \end{equation*}
    However, users may override this warning and proceed with unit-level modeling. All available data are used in model fitting, though extreme sparsity may lead to overly smoothed estimates, with the fitted surface imposing abnormally low subnational variation. Careful uncertainty assessment is critical in this scenario. This is a thorny issue, and is always an issue, with unit-level models in particular.
\end{itemize}

\begin{figure}[!ht]
    \centering
    \includegraphics[width=0.8\linewidth]{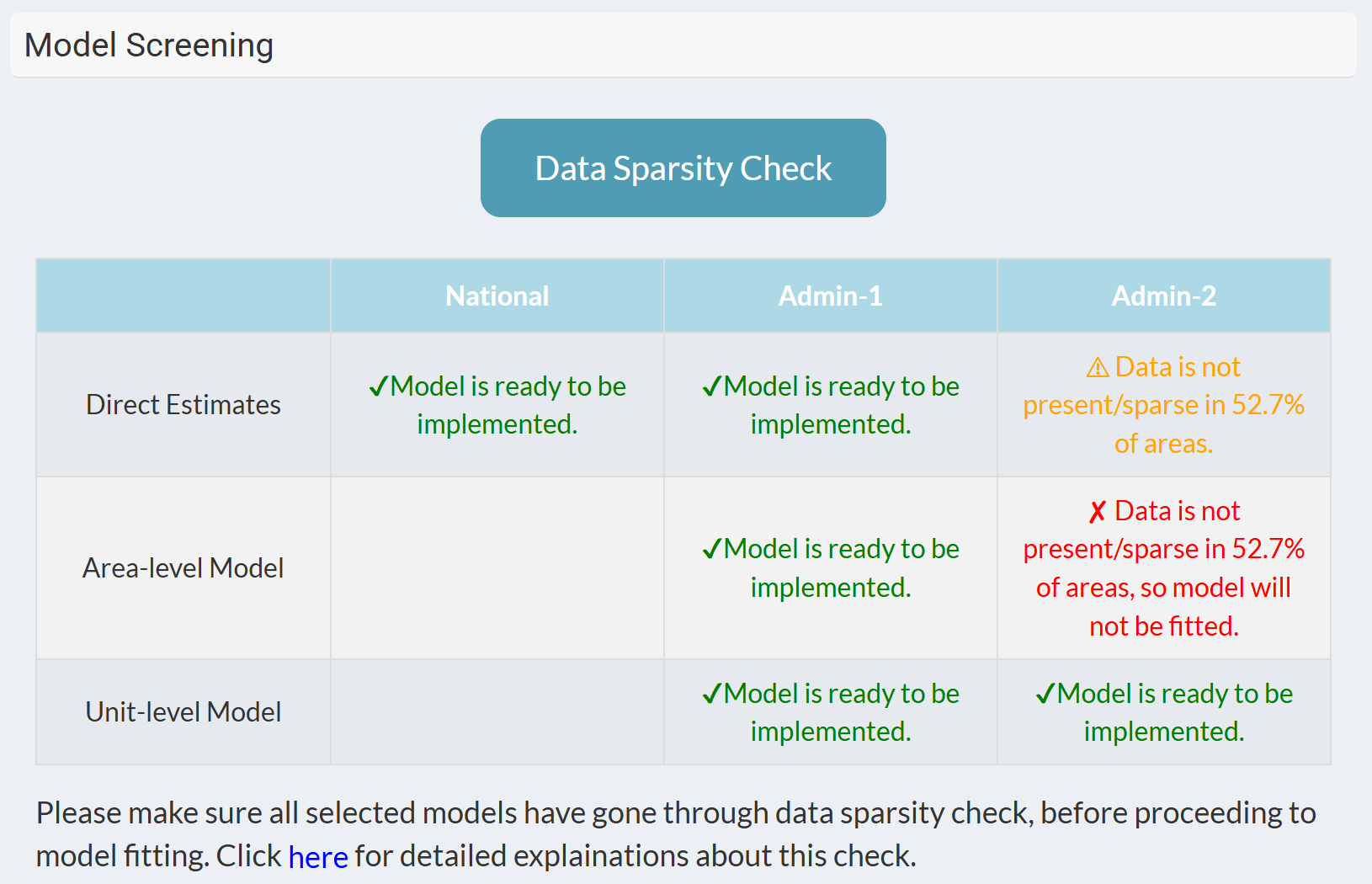}
    \caption{Recommendations for model fitting provided by \texttt{sae4health}}
    \label{fig:app_data_sparsity_check}
\end{figure}

Figure \ref{fig:app_data_sparsity_check} presents a screenshot of the warning messages in the model fitting interface, alerting users to potential data sparsity issues. 

\section{Case Study: Stunting for Under-five Children in Nigeria}

This section demonstrates \texttt{sae4health}'s features and functionality through a worked example, illustrating key modules such as data upload, model selection and fitting, visualization, and automated result reporting. Using data from the 2018 Nigeria DHS, we estimate the stunting rate among children under five, where stunting is defined as height-for-age Z-score (HAZ) less than -2. Selected screenshots highlight essential steps, while additional examples, extended documentation, and demo videos are available in the supplemental material and at \url{https://sae4health.stat.uw.edu}.

To launch \texttt{sae4health}, users can either access the web-based interface hosted on the remote server (\url{https://rsc.stat.washington.edu/sae4health}) or run the application locally using the following R command:

\noindent \code{> \, library("sae4health")}\\
\code{> \, options(shiny.launch.browser = TRUE) \# optional}\\
\code{> \, run\_app() }

\noindent
Launching \texttt{sae4health} in the default browser ensures that external links directing to documentation can open correctly and provides a better user experience for interactive features.

\begin{figure}[ht]
    \centering
    \includegraphics[width=1\linewidth]{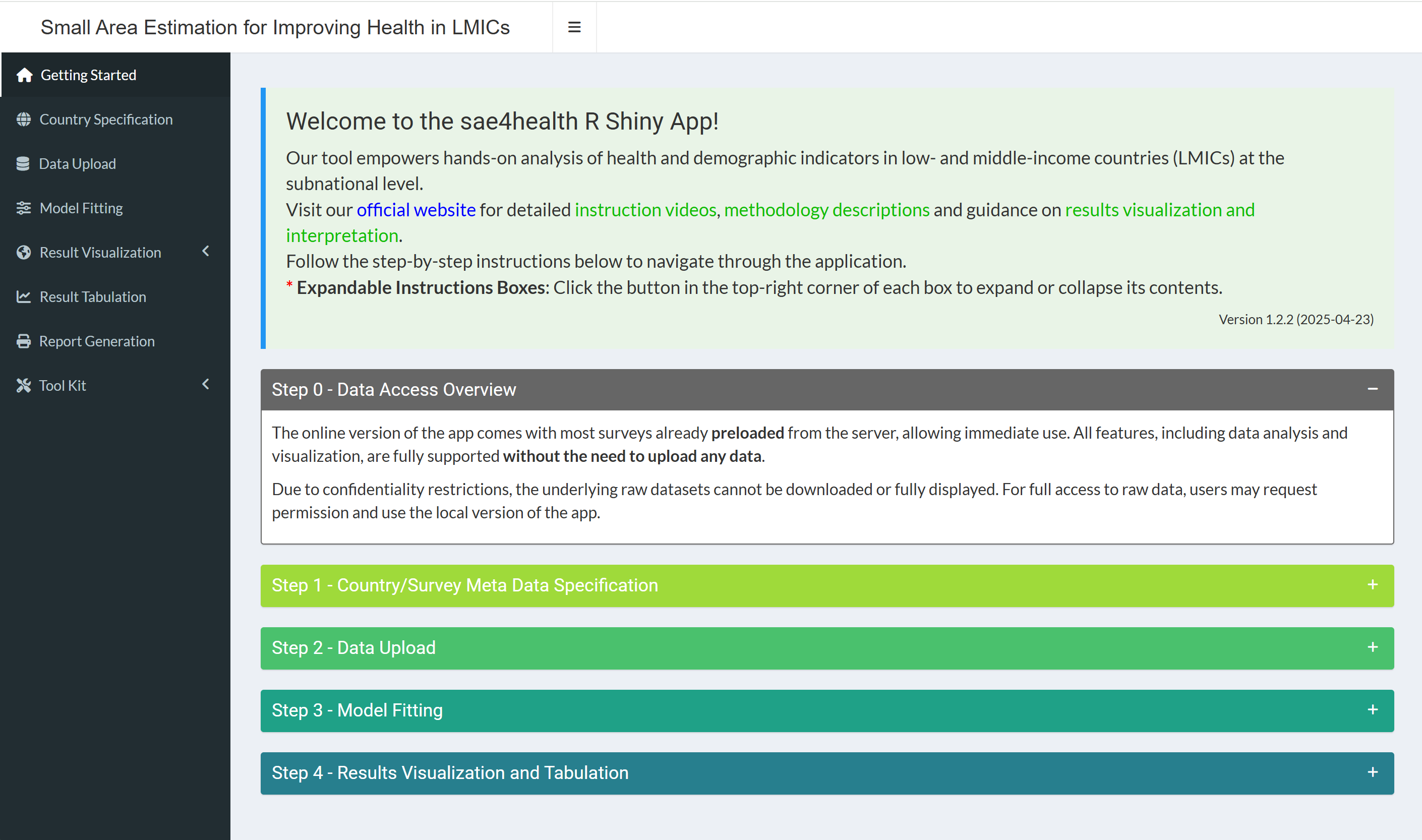}
    \caption{Landing page for the \texttt{sae4health} app, illustrating the workflow for SAE analysis.}
    \label{fig:app_landing_page}
\end{figure}

\subsection*{Landing Page}

The application opens on a landing page that provides a structured workflow for navigating its features. As depicted in Figure \ref{fig:app_landing_page}, the menu on the left outlines the key steps which include: \textbf{Country Specification}, \textbf{Data Upload}, \textbf{Model Fitting} and \textbf{Result Visualization}. This intuitive interface is designed to streamline the analysis of subnational health indicators. On this page, the app also offers comprehensive, step-by-step guidance for each module, as depicted in the expandable boxes.

\subsection*{Country Specification}
\begin{figure}[!ht]
    \centering
    \includegraphics[width=1\linewidth]{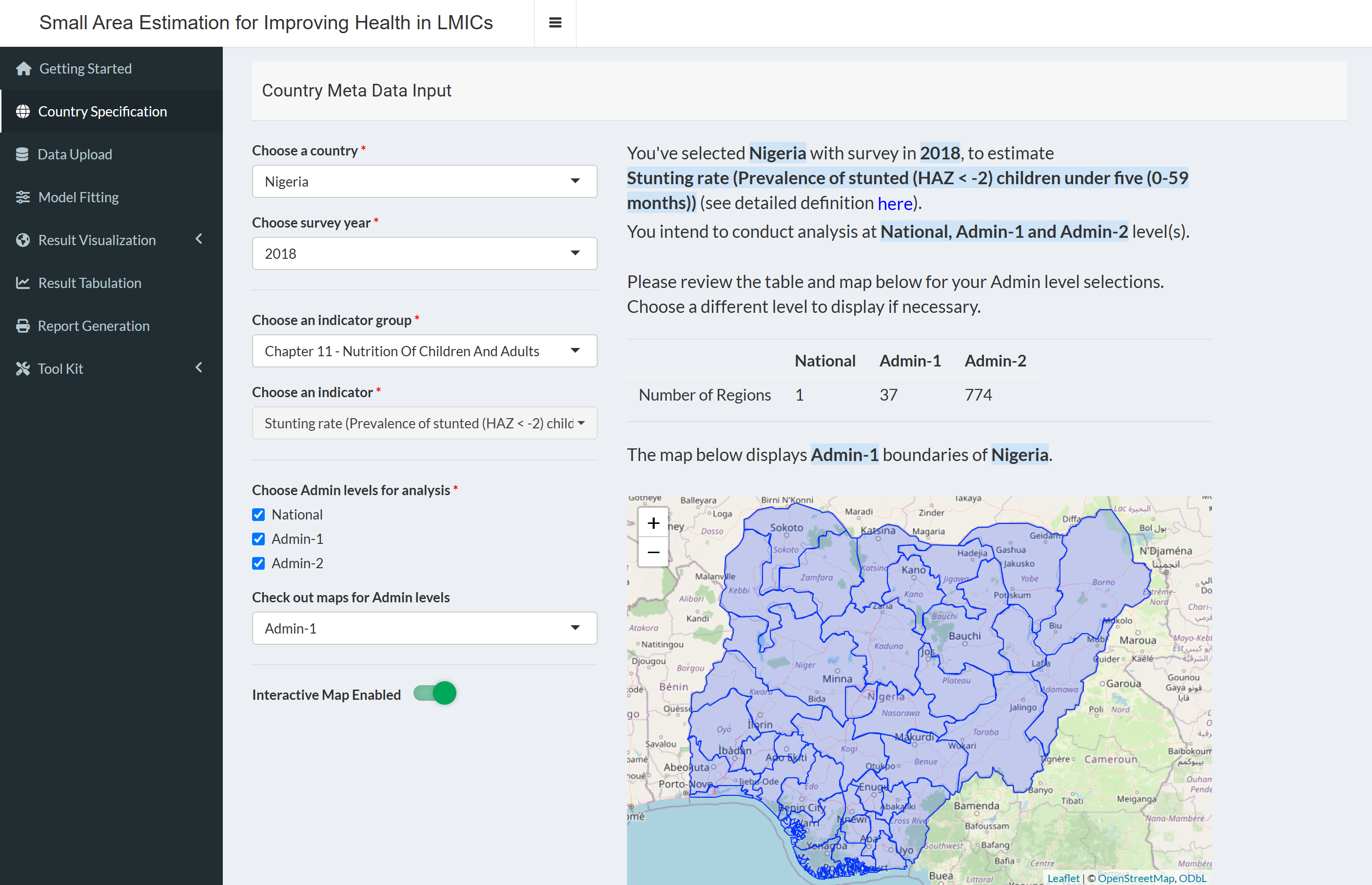}
    \caption{Country specification screen showing country boundaries and regional map.}
    \label{fig:app_country_specification}
\end{figure}

The first step is to specify the country, survey year, and indicator for analysis, as shown in Figure \ref{fig:app_country_specification}. Indicators are organized according to the chapters of the official DHS report, which group them by themes such as child health, family planning, and maternal health. This structure helps users efficiently locate relevant indicators. In this example, we focus on the stunting for under-five children indicator from the nutrition of children and adult category. To support geographic selection, administrative boundaries, sourced from the GADM database \citep{gadm}, are displayed as interactive maps. When a country and survey year are selected, the app automatically retrieves the corresponding boundary data using the R package \texttt{geodata} \citep{geodata} (preloaded for the web-based version). 

\subsection*{Data Loading}
In the next step, data are processed for analysis. In the web-based version, where datasets are available from the server, users simply click ``Load Data from Server`` to begin analysis without needing to obtain and manually upload files. For the local version of the app, users need to first request and upload their own DHS data. 
% Different health and demographic indicators rely on different DHS recode files (e.g., PR for household members, IR for women, BR for births). To support this, we provide detailed instructions for downloading the appropriate recode(s) along with the corresponding GPS cluster coordinates, which are essential for subnational mapping.

After clicking the button, \texttt{sae4health} reads, and processes the internally available (or uploaded) DHS files based on the coding scheme from the \texttt{surveyPrev} package. This may take a few moments depending on the file size. Once all required recode and GPS files are validated, the status icons in the upper-right corner turn green, and the cluster maps and modeling tools become available.

\begin{figure}[!ht]
    \centering
    \includegraphics[width=1\linewidth]{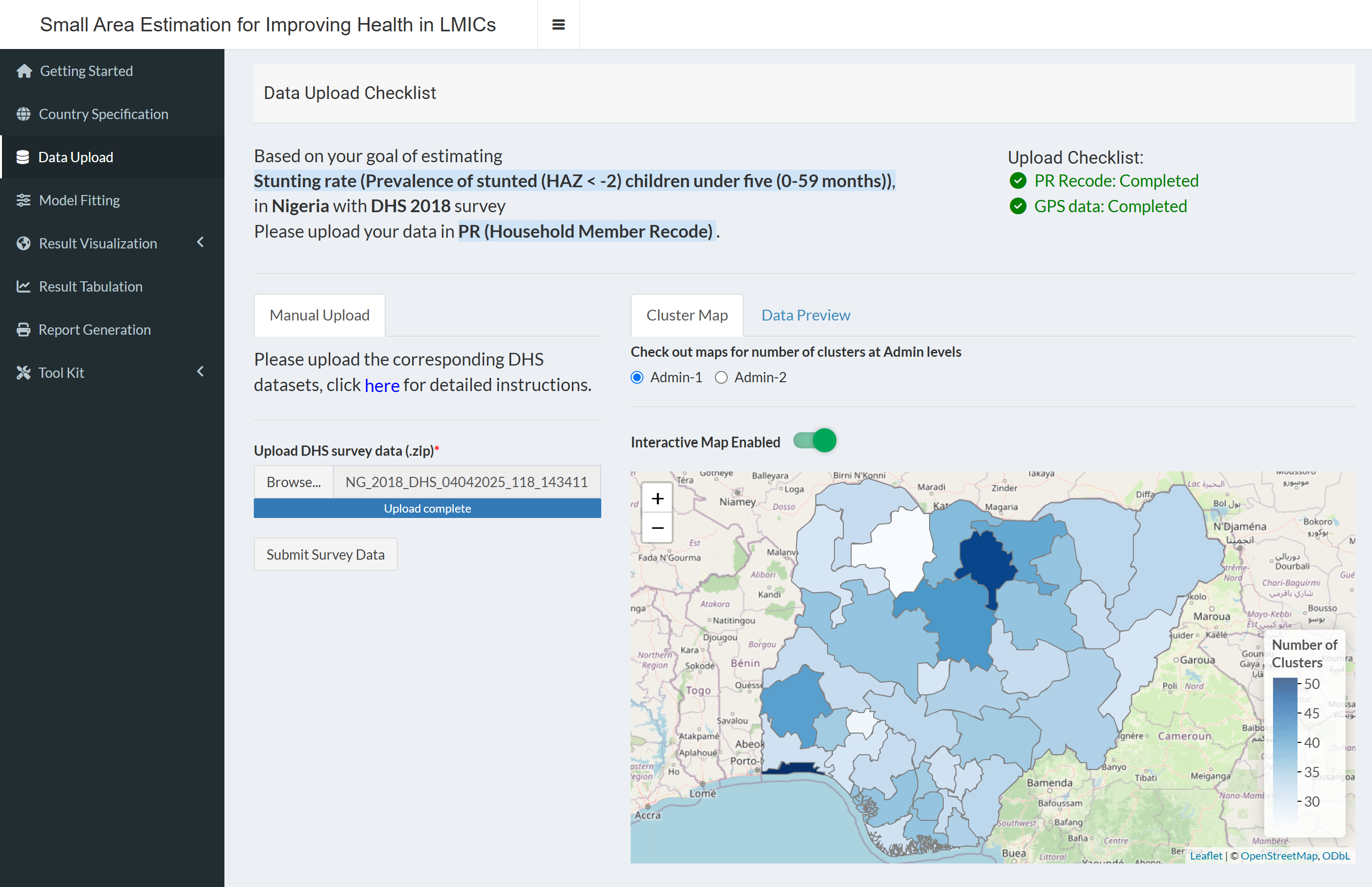}
    \caption{Data upload screen with a cluster map providing initial data sparsity diagnostics.}
    \label{fig:app_data_upload}
\end{figure}

As shown in Figure \ref{fig:app_data_upload}, diagnostic cluster map summarizes the number of surveyed clusters in each administrative unit. This initial visualization helps identify regions with sparse data, which is critical for guiding model selection. Here, and on all maps, one can hover over areas and additional information is retrieved. A ``Data Preview'' tab also allows users to inspect the structure of the raw analysis dataset.

\subsection*{Consistency of Estimates Check}
\begin{figure}[!ht]
    \centering
    \includegraphics[width=1\linewidth]{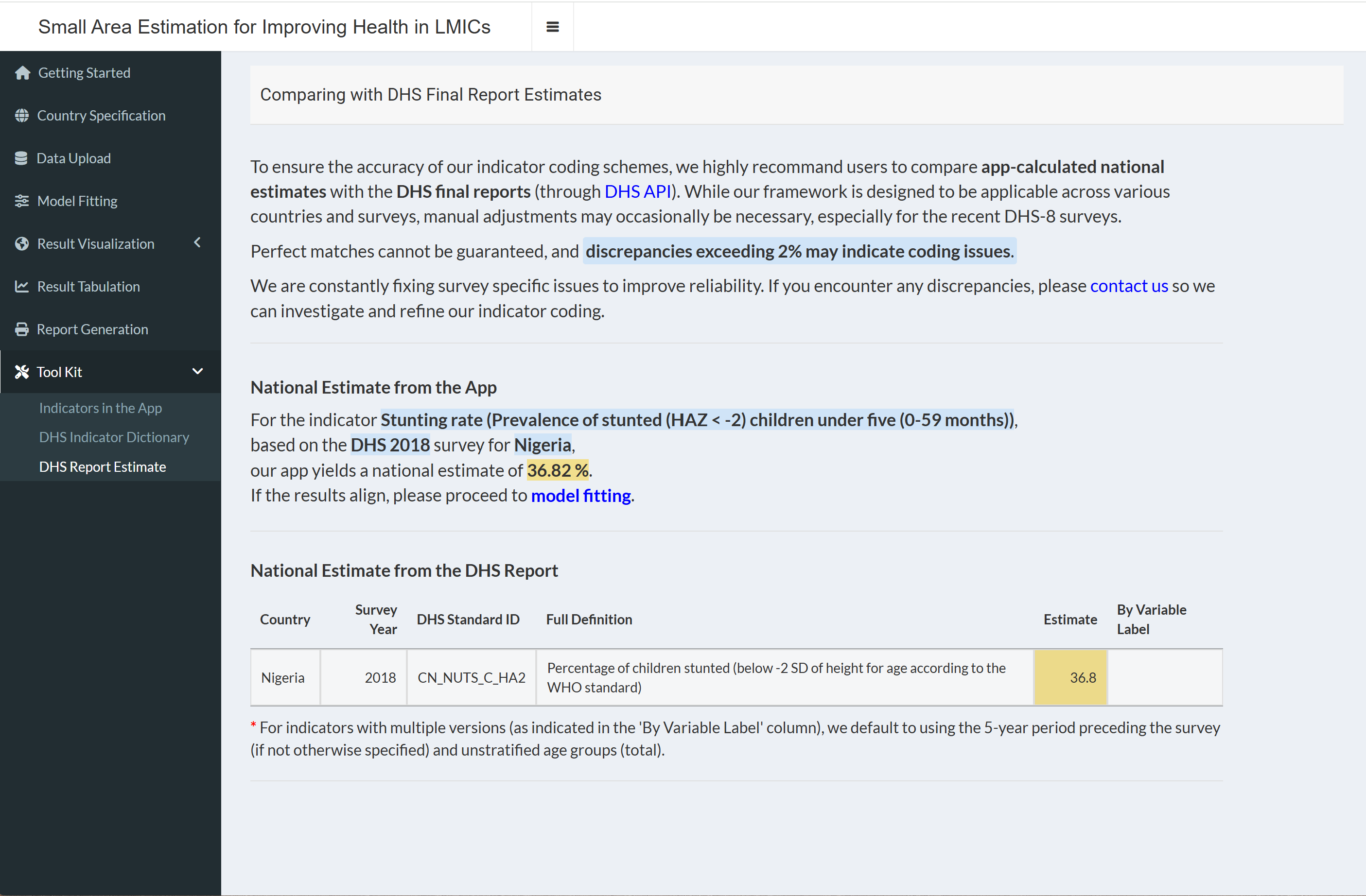}
    \caption{Comparing app-calculated estimates with official estimates.}
    \label{fig:app_estimate_check}
\end{figure}

To enhance the accuracy of indicator coding schemes, \texttt{sae4health} includes a feature for comparing app-generated national estimates with those reported in the DHS final reports. If discrepancies are detected, the app can alert users, prompting further review or adjustments to maintain consistency and reliability. The relevant page is depicted in Figure \ref{fig:app_estimate_check}.

\subsection*{Model Fitting}

The model-fitting process begins with a data sparsity check described before. If a model fails the data sparsity check, such as the area-level model at Admin-2, the app prevents fitting to ensure statistical validity. When sparsity may compromise interpretation, as with direct estimates at Admin-2, a warning is displayed.
 
\begin{figure}[!ht]
    \centering
    \includegraphics[width=1\linewidth]{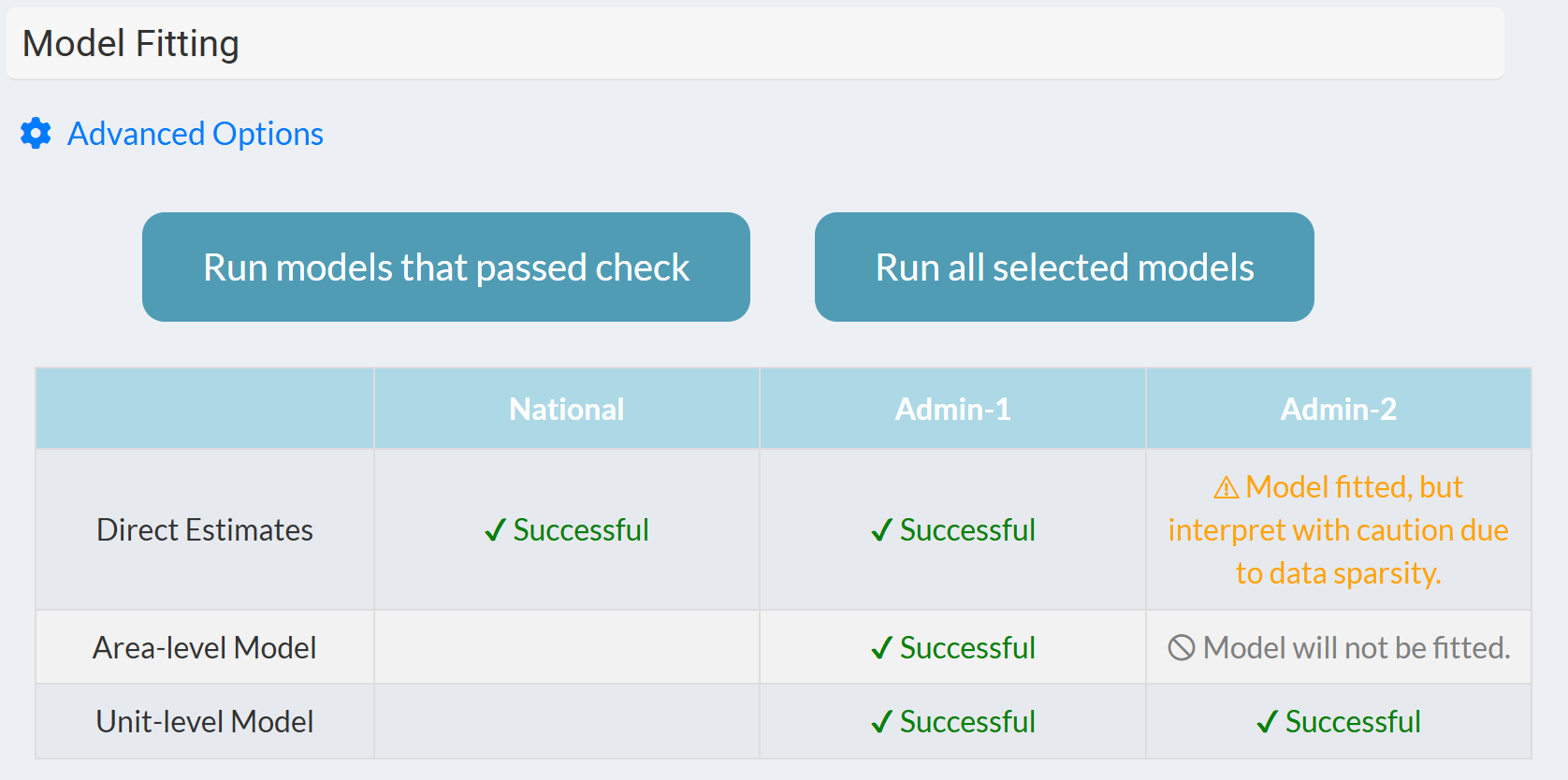}
    \caption{Interface for model fitting results.}
    \label{fig:app_model_fitting_3}
\end{figure}

Before selecting the models to fit, users can also customize the model using the ``Advanced Option'' panel. Currently, \texttt{sae4health} supports a `nested model' option, which adds Admin-1 fixed effects to Admin-2 models to mitigate over-smoothing, and also an `area-level covariate' option. For the case study presented (under-five stunting in Nigeria), we use the default settings, i.e., non-nested models and no covariates. Further details on the implementation of these advanced options are provided in section ``Advanced Options'' in the supplemental material. The user can proceed with model fitting by clicking the corresponding button(s), after which the fitting status is displayed, as shown in Figure \ref{fig:app_model_fitting_3}. 

\subsection*{Result Visualization}

\begin{figure}[!ht]
    \centering
    \includegraphics[width=1\linewidth]{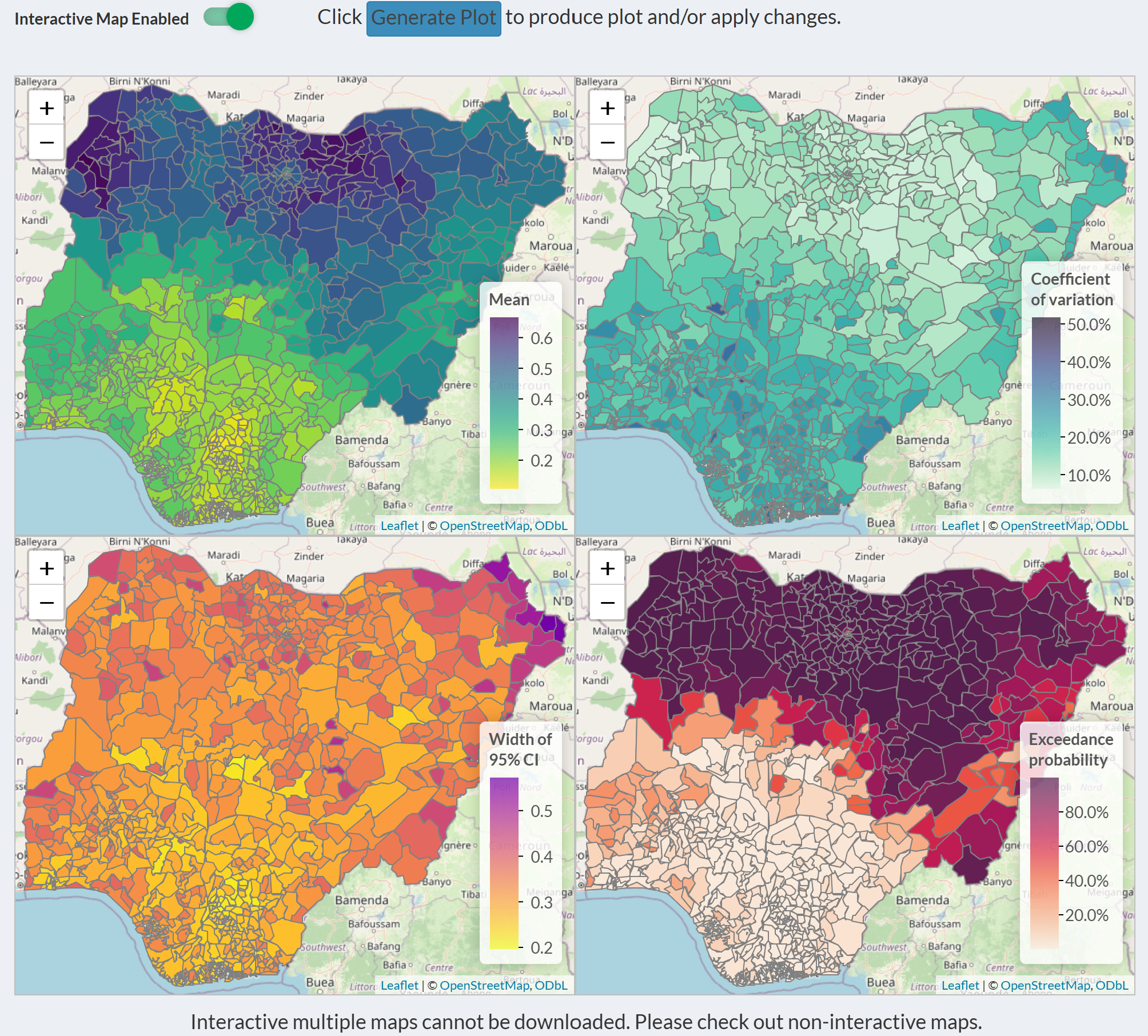}
    \caption{Maps showing key statistics across regions, including point estimates, widths of uncertainty intervals, coefficients of variation and exceedance probabilities.}
    \label{fig:app_prev_map_2}
\end{figure}

We offer a comprehensive suite of visualization tools in \texttt{sae4health} to support the exploration and interpretation of subnational prevalence estimates. Recognizing that point estimates alone are insufficient for robust decision-making, the app emphasizes uncertainty through multiple complementary displays. Users can interact with visualizations by hovering over regions for detailed statistics, or export high-quality static versions in PDF format for reporting and dissemination purposes. A simple toggle bar enables easy switch between interactive and static modes. The full suite of visualization tools is described below.

\noindent \textbf{Prevalence Maps}: Prevalence maps are essential for visualizing regional variation in health and demographic indicators. As illustrated in Figure~\ref{fig:app_prev_map_2}, \texttt{sae4health} displays key statistical measures, here for under-five stunting rates in Nigeria at the Admin-2 level. A clear spatial pattern emerges, with higher stunting rates concentrated in the northern regions and noticeable spatial disparities across subnational areas, with 8-fold variation in the point estimates.

Alongside point estimates, the app presents several uncertainty measures, such as the lengths of 95\% confidence intervals (credible intervals for model-based estimates). In Figure~\ref{fig:app_prev_map_2} we see that these widths are very wide, reflecting the paucity of information at Admin-2. To further assess reliability, the app visualizes the coefficient of variation (CV), defined as
\[
\text{CV} = 100 \times \frac{\text{Standard Error}}{\text{Point Estimate}},
\]
which provides a standardized measure of relative uncertainty for the weighted estimates and with the posterior standard deviation replacing the Standard Error for the Bayesian summaries (area-level and unit-level models).

In addition, the app supports exceedance probability mapping, which shows the probability that prevalence in a given area exceeds a user-defined threshold \( p_0 \), calculated as
\[
\Pr(p_i > p_0 \mid y).
\]
This feature is especially useful for identifying regions that exceed a pre-specified benchmark, helping to flag high-burden areas for targeted interventions. In Figure~\ref{fig:app_prev_map_2} we plot the exceedence probabilities with $p_0$ taken as the national average of 0.37, and see a stark contrast between the north and the south.

\begin{figure}[!ht]
    \centering
    \includegraphics[width=1\linewidth]{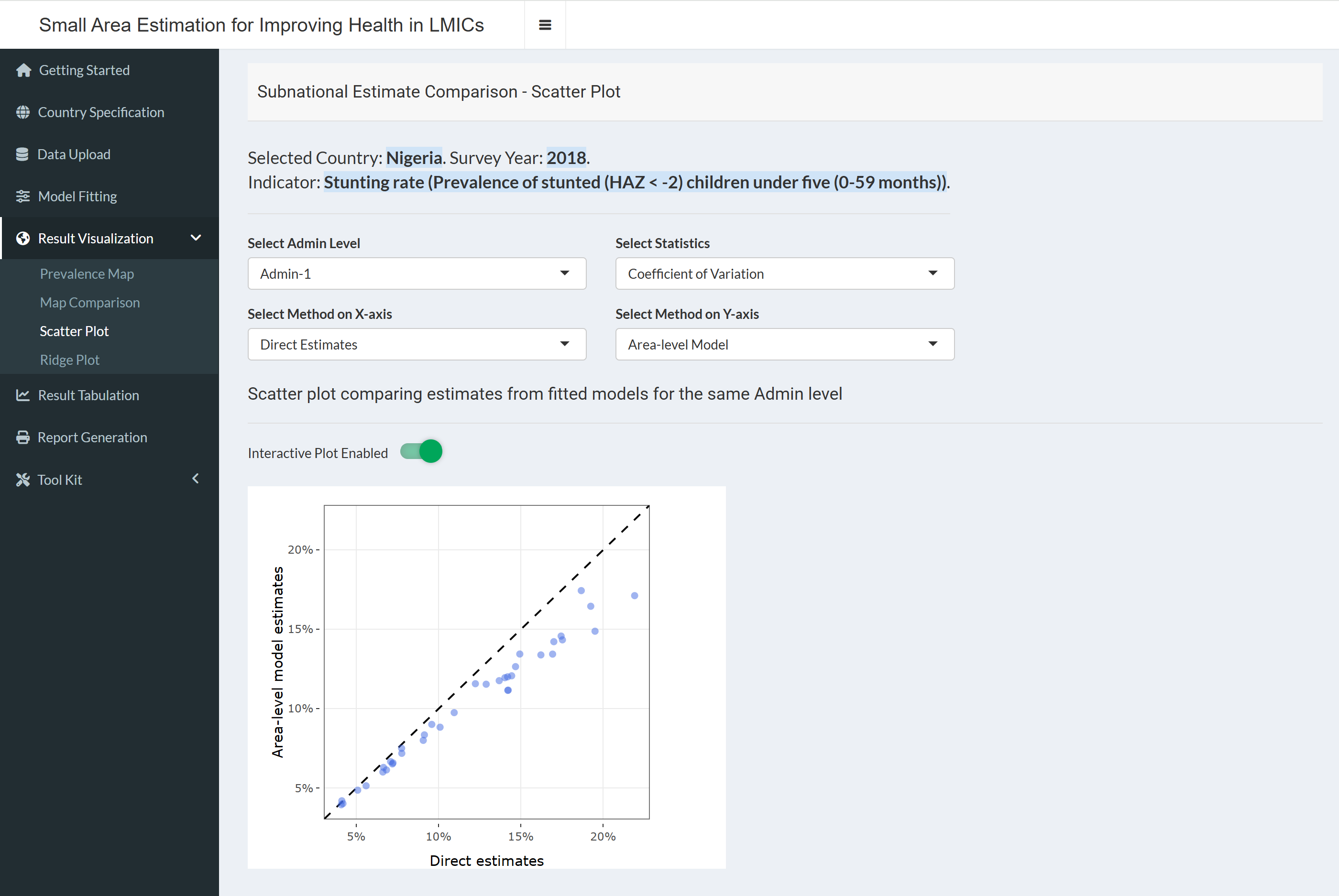}
    \caption{Scatter plot comparing point estimates from direct and model-based estimates.}
    \label{fig:app_scatter_plot}
\end{figure}
\noindent \textbf{Scatter Plot:} Figure~\ref{fig:app_scatter_plot} illustrates the scatter plot feature presented by \texttt{sae4health}, which compares univariate summaries from two different modeling approaches. For instance, in this example, we compare the CV of under-five stunting in Nigeria across Admin-1 regions. The $x$-axis represents the CV of the direct estimates, while the $y$-axis represents those from the area-level model.

This plot is valuable for evaluating the effect of smoothing in model-based approaches. Model-based estimates often exhibit improved precision, as they borrow strength from neighboring areas. This behavior is evident with points below  the diagonal line (lower CV from the area-level model). By comparing point estimates between models, the scatter plot is useful in assessing the trade-off between precision and potential bias introduced by smoothing.

\begin{figure}[!ht]
    \centering
    \includegraphics[width=1\linewidth]{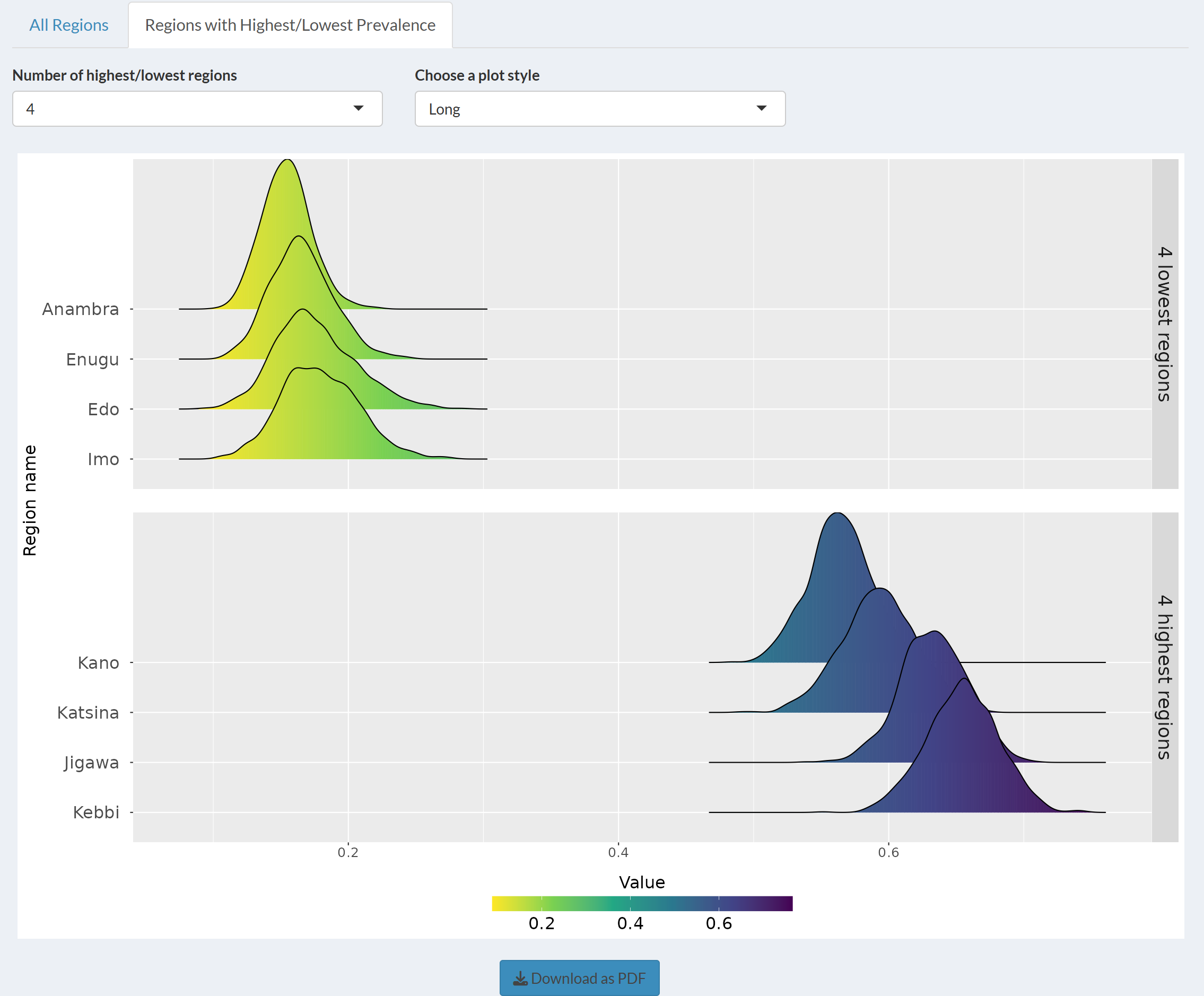}
    \caption{Ridge plot showing the posterior distributions of prevalence estimates for extreme regions.}
    \label{fig:app_ridge_plot}
\end{figure}

\noindent \textbf{Ridge Plot:} Ridge plots, shown in Figure~\ref{fig:app_ridge_plot}, offer a comprehensive view of the full (posterior) distribution of prevalence estimates across selected regions. These visualizations serve as valuable inferential tools for identifying areas that
require special attention.

Overlapping distributions suggest that differences between regions may not be statistically significant, whereas clearly separated ridges indicate substantive disparities. This type of visualization is especially informative for ranking or prioritizing regions based on relative prevalence, while accounting for the uncertainty inherent in the estimates. 

\texttt{sae4health} provides flexible options for generating ridge plots. Users can choose to display all Admin-1 regions or focus on Admin-2 regions within any specified Admin-1. Alternatively, the app can display the top \( x \) regions with the highest and lowest prevalence values across the country. We have found these plots pedagogically very useful, since they show the inherent variation for specific areas, while maps show the between-area variation. Ridege plots offer a sobering picture of the often considerable uncertainty, even at Admin-1.

\subsection*{Result Tabulation}
\begin{figure}[!ht]
    \centering
    \includegraphics[width=1\linewidth]{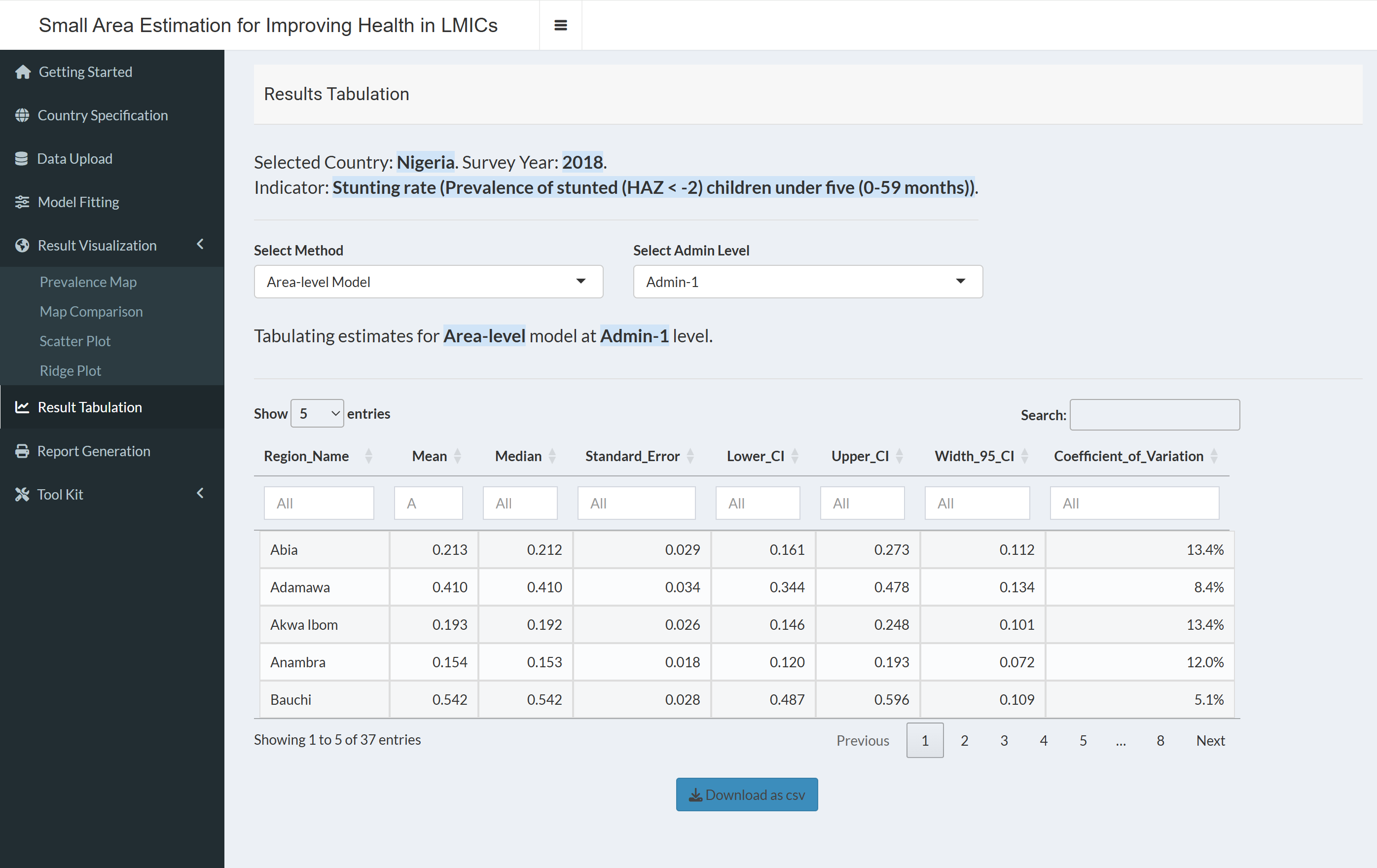}
    \caption{``Result Tabulation'' tab showing tabulated estimates with options to download as a CSV file.}
    \label{fig:app_result_tabulation}
\end{figure}

The “Result Tabulation” tab enables users to review exact estimates for each location, method, and Admin level. As shown in Figure \ref{fig:app_result_tabulation}, users can adjust the number of rows displayed and navigate between pages if needed. Additionally, a download option is provided, allowing users to export the tabulated estimates as a .csv file for further analysis.

\subsection*{Report Generation}
The ``Report Generation'' tab provides an automated way to compile results and visualizations into a comprehensive report. This feature allows users to easily document their analysis and share findings with stakeholders. The generated report includes relevant figures, tables, and summaries from the conducted analysis. For reference, a sample report automated by the app is included in the supplemental material.

\section{Discussion}
\label{sec:discussion}

The \texttt{sae4health} R Shiny application demonstrates how statistically rigorous SAE analyses can be operationalized in an accessible tool for routine use in data-sparse LMICs settings. The app enables users to generate subnational estimates for over 150 health and demographic indicators derived from DHS surveys under different models. The complete data analysis workflow is integrated into a unified interface. Users can load DHS data, define model settings, conduct design-based and model-based SAE analysis, and visualize results through interactive outputs.  The graphical user interface eliminated many technical barriers and allows the methods to be accessible to users without requiring programming skills or advanced statistical training. Model choice and uncertainty quantification are emphasized throughout the workflow. It is particularly suited for public health researchers and officials in low-resource settings.

While the current implementation of \texttt{sae4health} supports a wide range of indicators and flexible modeling options, ongoing development aims to further enhance customization, expand data source compatibility, and support additional features to meet evolving analytic needs. Increasing the suite of model validation tools is a priority.

One major expansion is the integration of Multiple Indicator Cluster Surveys (MICS), a widely used household survey in LMICs with the same design as DHS. While the methodological framework for SAE can be readily adapted to MICS data, the challenge lies in its non-standardized indicator coding scheme and less harmonized data structures across countries, requiring additional processing to prepare analysis dataset. However, this challenge is surmountable, such that we have a Nigeria MICS app available, and aim to develop a standardized pipeline for integrating more MICS surveys into \texttt{sae4health}. The MICS app can be reached from \url{https://sae4health.stat.uw.edu/app_MICS}.

Another critical improvement focuses on addressing systematic bias in unit-level models. The current implementation in \texttt{sae4health} does not fully acknowledge the survey design, which can lead to significant biases in prevalence estimates \citep{wu2024modelling,dong2021modeling}. This bias primarily stems from urban-rural oversampling in DHS surveys and the differences in estimates for the indicator of interest between urban and rural areas. To mitigate this, we have to fit separate unit-level models for urban and rural data, and aggregate urban/rural specific estimates using urban-rural population weights to obtain the overall estimates. The methodology outlined in \cite{wu2024modelling} offers a foundation for systematically resolve such bias, but further adaptation is required to integrate the pipeline seamlessly into our existing workflow.

We also plan to expand \texttt{sae4health} by incorporating composite indicators such as the under-five mortality rate (U5MR) and fertility rates. The DHS spatial report by \cite{wu:etal:21} provides a structured approach for estimating U5MR using spatio-temporal SAE models at the Admin-2 level based on DHS data. While the pipeline is conceptually aligned with our framework, its modeling structure differs significantly (because we need to model mortality as a function of age, which increases model complexity), requiring additional work on result interpretation and model diagnostics.

Beyond the above improvements, we plan to incorporate additional visualization tools, such as ranking plots, to support more robust statistical inference. Additionally, we aim to incorporate continuous spatial models for more precise geographic estimation and explore the integration of auxiliary unit-level covariates to improve model predictions, to supplement the currently available inclusion of area-level covariates.

Last but not least, development of \texttt{sae4health} has been shaped by feedback from users and researchers in sub-Saharan African countries, through direct collaboration and training workshops organized by the World Health Organization (WHO) and UNICEF. We summarize some of these experiences on the ``Impact'' tab of \url{https://sae4health.stat.uw.edu}. From our experience, it is clear that there is a crucial gap between academic research and tools that are accessible to researchers and policymakers in LMICs. Beyond providing a graphical user interface, a more important role for \texttt{sae4health} is to promote a rigorous and reproducible workflow for obtaining reliable estimates for key indicators that is adapted to local contexts, without needing to manage complex data structures or implement sophisticated modeling procedures from scratch. We aim for \texttt{sae4health} to serve as a practical tool for data-driven public health planning, monitoring, and decision-making in LMICs.

\newpage

\section{Supplementary Material}
\label{sec-supp}

\subsection{Advanced Options}
\label{sec-supp-ad-options}
\begin{figure}[!ht]
    \centering
    \includegraphics[width=0.85\linewidth]{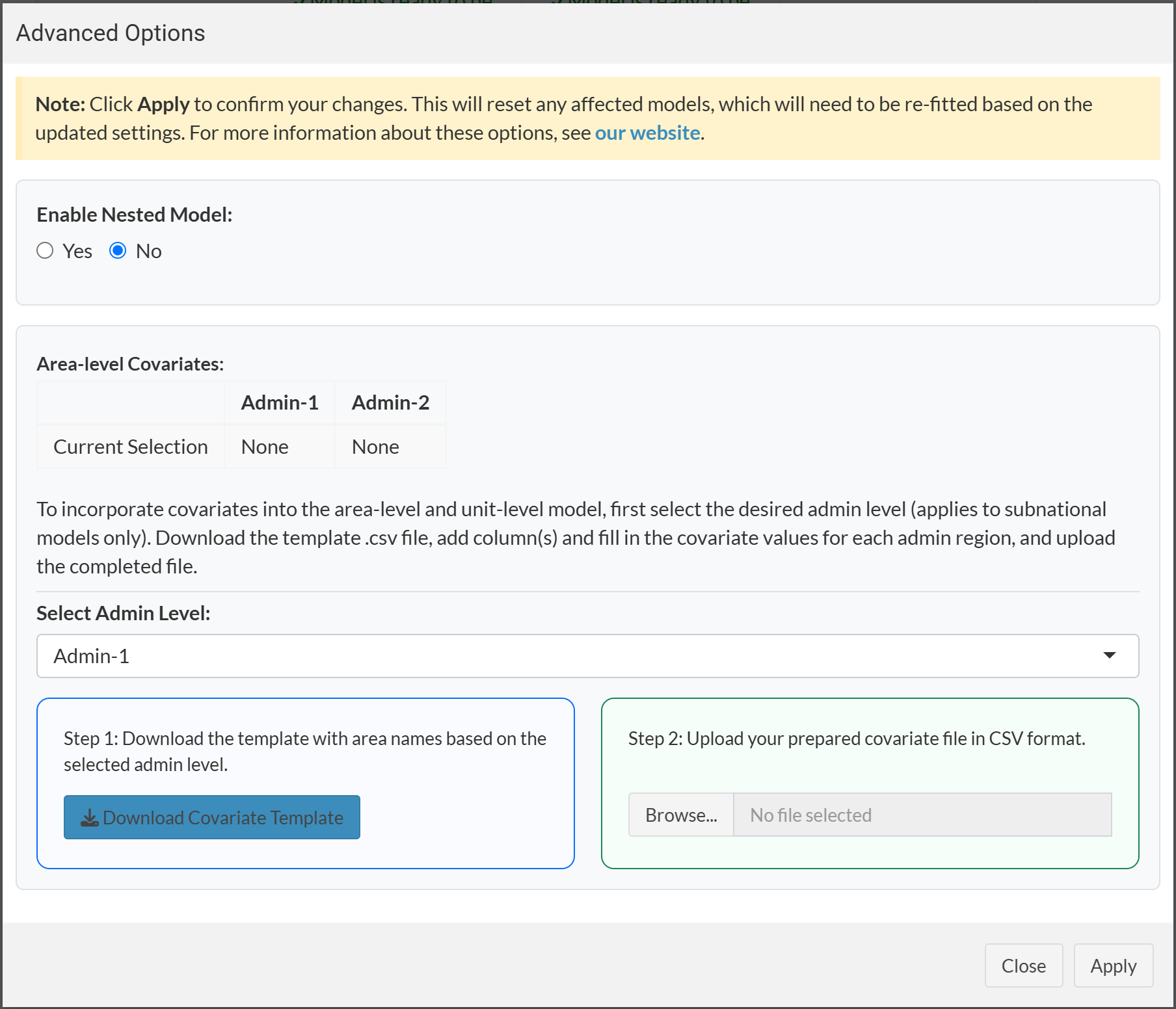}
    \caption{Advanced option panel with default settings: nested-model disabled and no covariates.}
    \label{fig:ad_op_before}
\end{figure}

After completing the data sparsity check, users have the option to customize model specifications through the   ``Advanced Options'' panel. By expanding this panel (see Figure \ref{fig:ad_op_before}), users can enable additional modeling features, such as the nested model and covariate incorporation.

\noindent \textbf{Nested Admin-2 Model}:  The spatial models implemented in \texttt{sae4health} smooth the raw survey estimates both locally and nationally. While smoothing improves estimation accuracy in areas with limited data, \textbf{oversmoothing} can become a concern, particularly when data are sparse. Recall that surveys are powered to produce reliable estimates for some indicators at Admin-1 and, hence, we would like to recover the weighted estimates, rather than distort away from those.

To address this, \texttt{sae4health} offers a `nested model' option. In this approach, Admin-1 fixed effects are incorporated into the model:
\begin{equation*}
p_{ic} = \text{expit} \left( \alpha_{a_c} + b_i\right),
\end{equation*}
where $a_c$ denotes the Admin-1 area containing cluster $c$, and \( \alpha_a \) are a set of Admin-1 level fixed effects which are included to effectively recover the weighted estimates. The terms $b_i$ then smooth variation \textbf{within} each Admin-1 area. This modeling structure can be viewed as being more consistent with the sampling scheme in which the Admin-1 areas are sampling strata.

Currently, the nested model option applies only to unit-level models at or below the Admin-2 level. By default, this option is disabled.\\

\noindent \textbf{Covariate Incorporation}: Users may incorporate \textbf{area-level covariates} into either area-level or unit-level models to improve model performance (default is no covariates). 

To do so, users first select the target administrative level and download the provided \texttt{.csv} template listing the corresponding region names. Covariate columns can then be added and populated with values for each region. After uploading the completed file back into the app, users must click the \texttt{Apply} button to confirm the changes. 

\begin{figure}[H]
    \centering
    \includegraphics[width=0.85\linewidth]{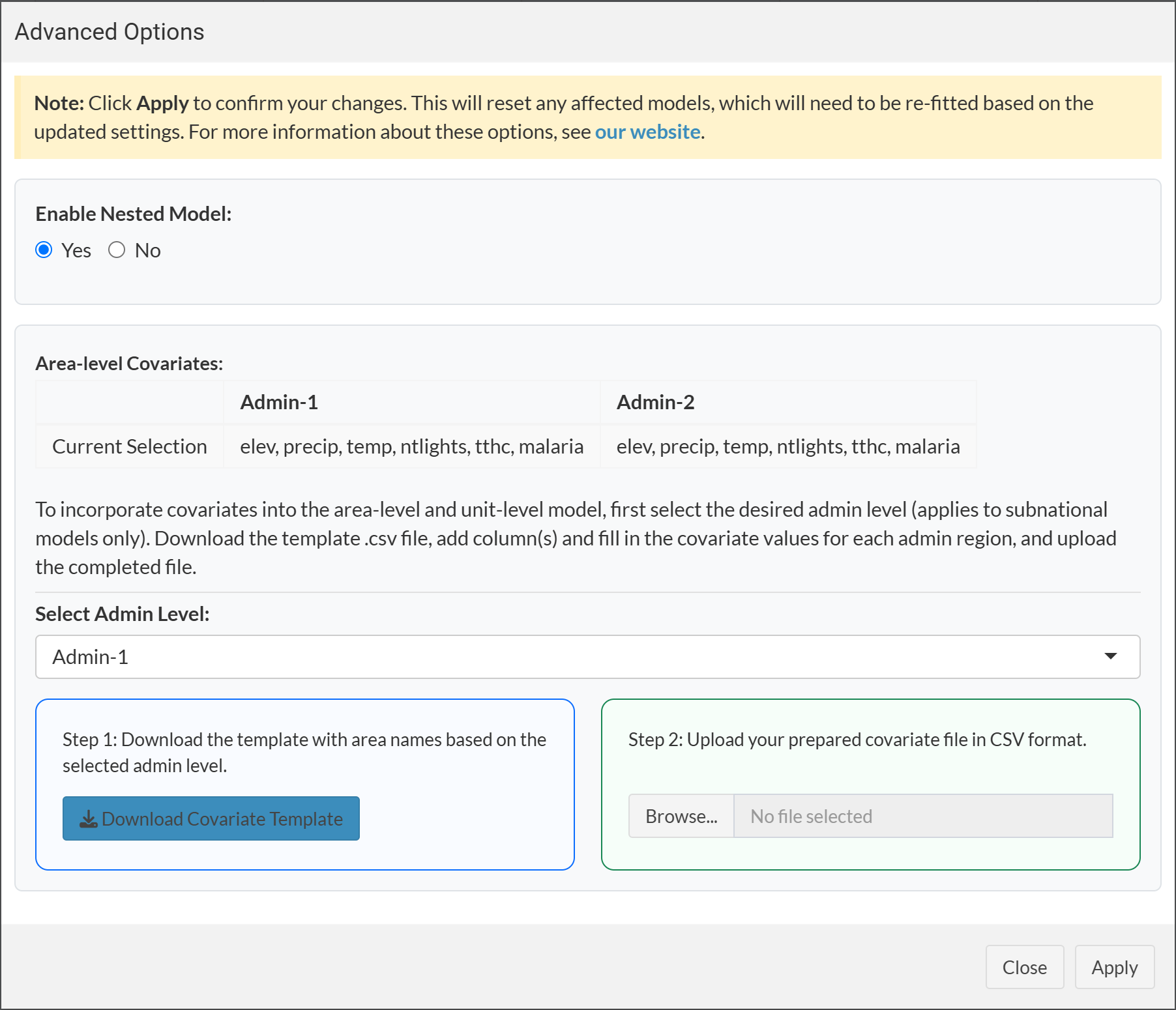}
    \caption{Advanced option panel showing names for successfully uploaded covariates.}
    \label{fig:ad_op_after}
\end{figure}

The status of the uploaded covariates can be verified by reopening the ``Advanced Options'' panel, where the names of the incorporated covariates will be displayed, as shown in Figure \ref{fig:ad_op_after}. Successfully uploaded covariates are automatically included in the model fitting process.

\newpage
\subsection{Sample Report}
\label{sec-supp-report}

After completing the workflow up to model fitting, users can generate a comprehensive, downloadable report with a single click within the app. This report includes metadata, sample details, data sparsity insights, and key visualizations such as prevalence maps and model comparison plots. Designed to streamline result sharing and interpretation, the report ensures that all relevant outputs are neatly compiled for reference and communication.

This section presents a sample report based on the analysis of stunting rate among children under five using data from the 2018 Nigeria DHS. This example demonstrates the report’s format, structure, and the range of insights it provides.



\section*{Supplementary material}

# Sample Report from the sae4health R shiny App

## Summary Info

Country: **Nigeria**

Survey: **2018**

Indicator: **Stunting rate (Prevalence of stunted (HAZ < −2) children under five (0–59 months))**

Levels: **National, Admin−1 and Admin−2**

Number of regions at selected admin levels:

|  | National | Admin−1 | Admin−2 |
|---|---|---|---|
| *Number of Regions* | 1 | 37 | 774 |

Detailed information on the indicator:

| DHS Standard ID | DHS Report Chapter | Definition |
|---|---|---|
| CN_NUTS_C_HA2 | Chapter 11 – Nutrition Of Children And Adults | Percentage of children stunted (below −2 SD of height for age according to the WHO standard) |

\* For indicators with multiple versions, the app defaults to the 5−year period before the survey (unless specified otherwise), and unstratified age groups (total).

# Sample Report from the sae4health R shiny App

## Estimate Consistency Check (National Level)

To ensure the accuracy of our indicator coding schemes, we highly recommend users to compare

**app–calculated national estimates** with the **DHS final reports**.

National estimate (from the app): **36.82 %**

National estimate (from DHS Final Report):

| Country | Survey Year | DHS Standard ID | Full Definition | Estimate | By Variable Label |
|---------|-------------|-----------------|-----------------|----------|-------------------|
| Nigeria | 2018 | CN_NUTS_C_HA2 | Percentage of children stunted (below –2 SD of height for age according to the WHO standard) | 36.8 | |

*For indicators with multiple versions, the app defaults to the 5–year period before the survey (unless specified otherwise), and unstratified age groups (total).

# Sample Report from the sae4health R shiny App

## Overall Sample Info

Total number of clusters: **1382**

Total sample size: **12381**

Total number of events: **4504**

Number of regions without any data:

|  | National | Admin–1 | Admin–2 |
|---|---|---|---|
| *Regions without Any Clusters* | 0% | 0% | 18.22% |

3

# Sample Report from the sae4health R shiny App

Sample Info for Admin−1

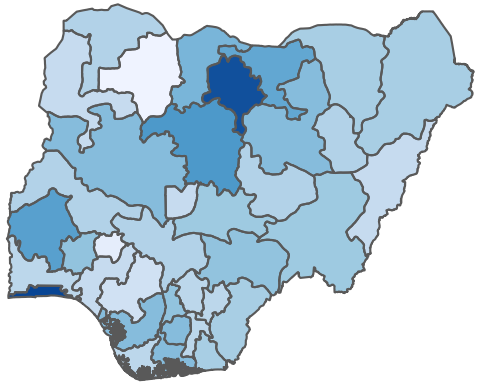

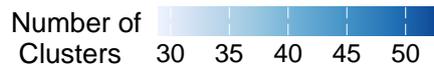

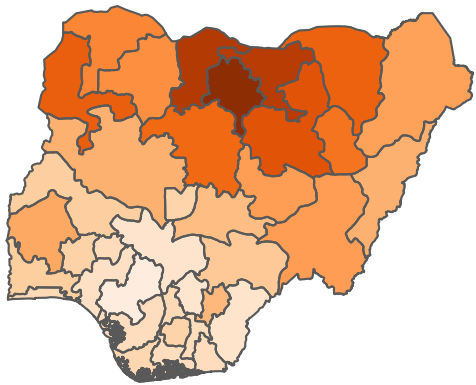

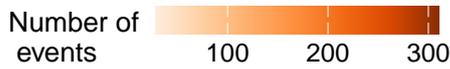

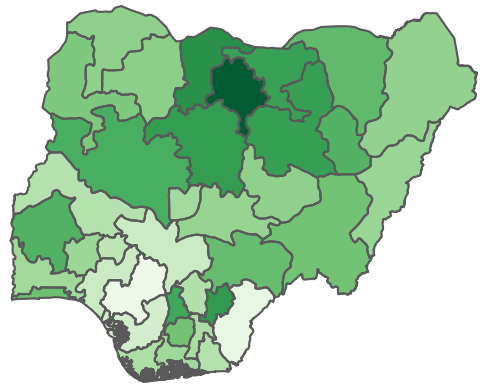

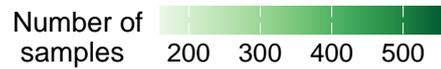

4

# Sample Report from the sae4health R shiny App

Sample Info for Admin−2

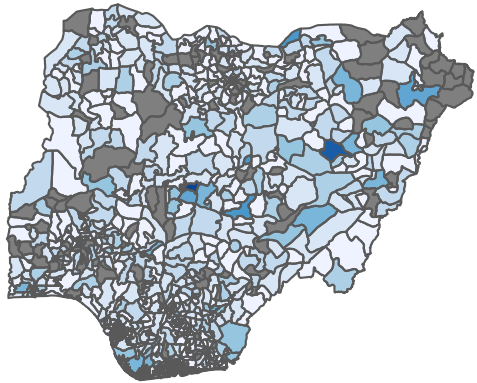
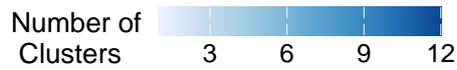

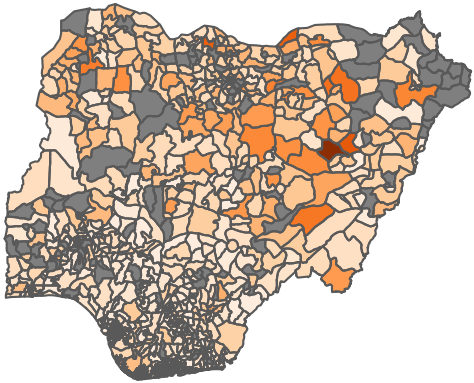
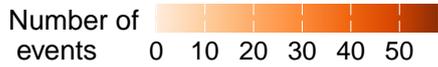

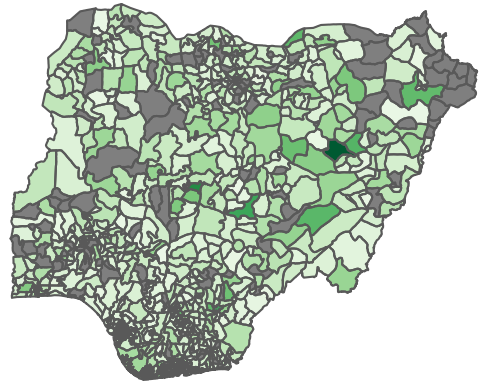
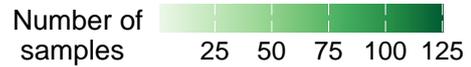

5

# Sample Report from the sae4health R shiny App

Key Statistics Maps: Direct estimates model at Admin−1 level

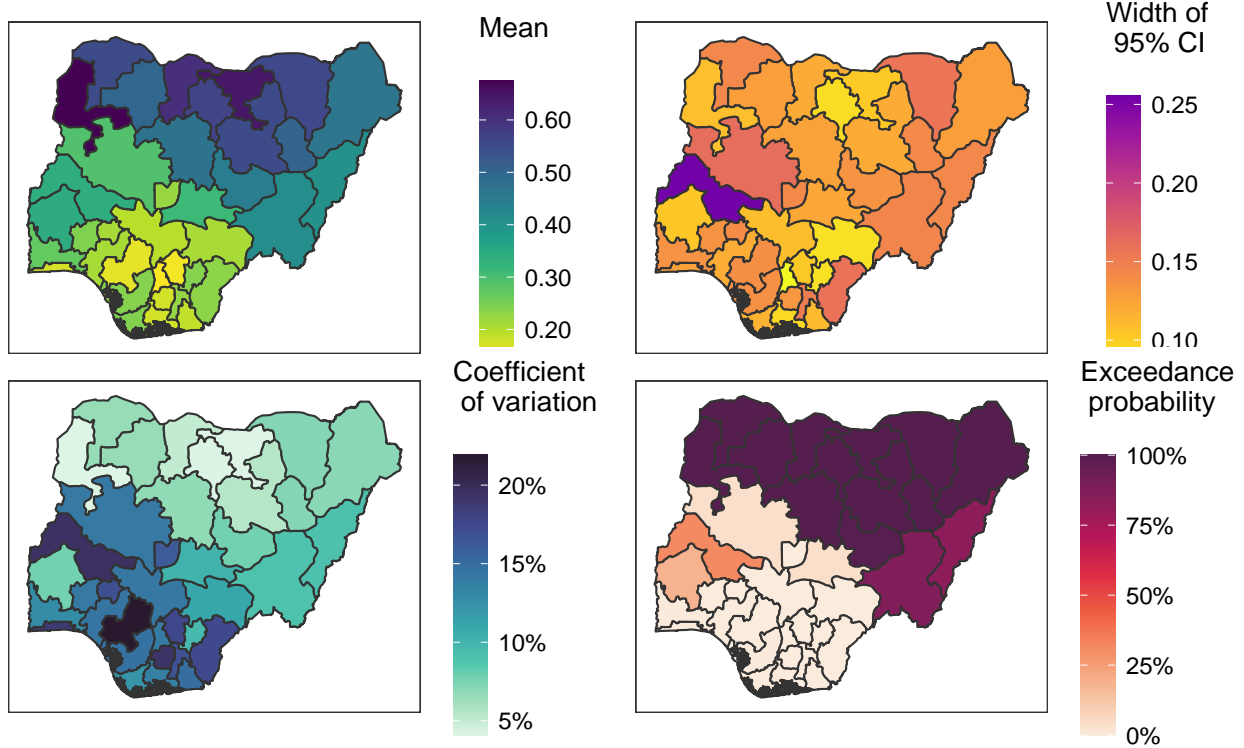

6

# Sample Report from the sae4health R shiny App

Key Statistics Maps: Direct estimates model at Admin−2 level

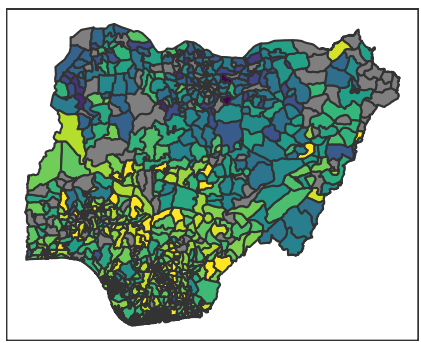 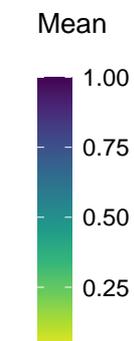 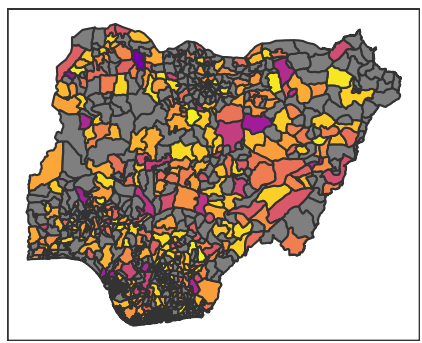 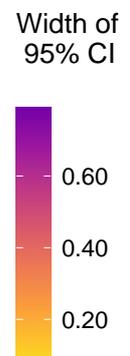

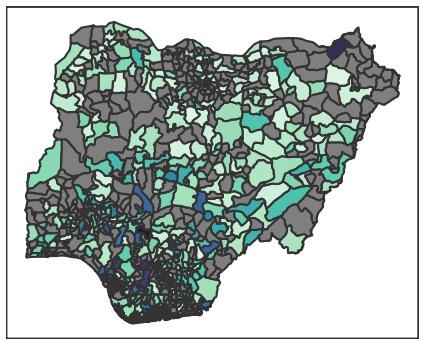 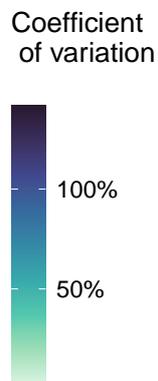 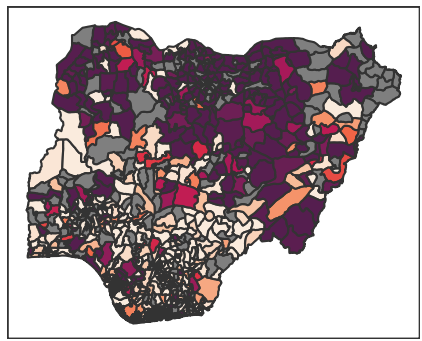 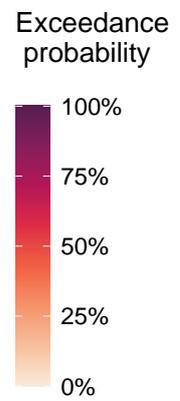

7

# Sample Report from the sae4health R shiny App

Key Statistics Maps: Area–level model at Admin–1 level

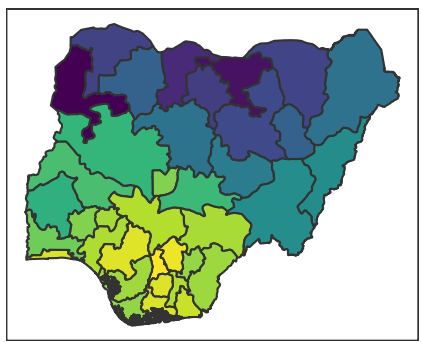 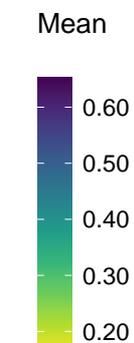 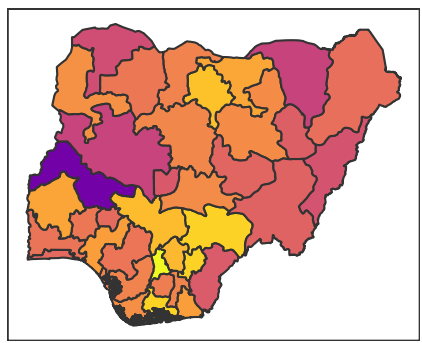 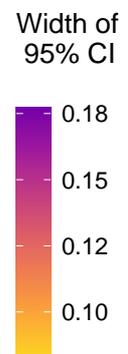

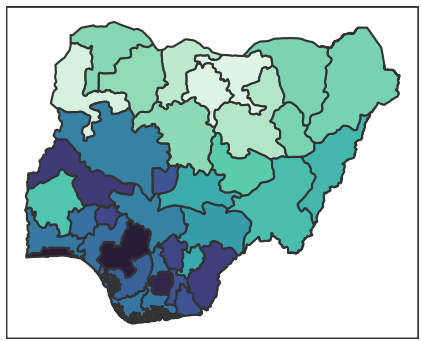 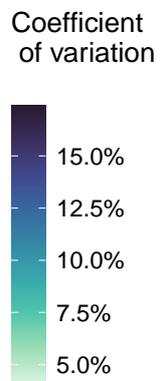 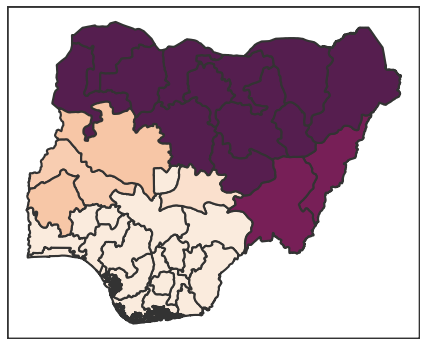 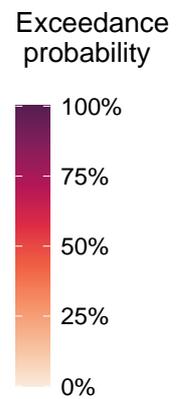

8

# Sample Report from the sae4health R shiny App

Key Statistics Maps: Unit−level model at Admin−1 level

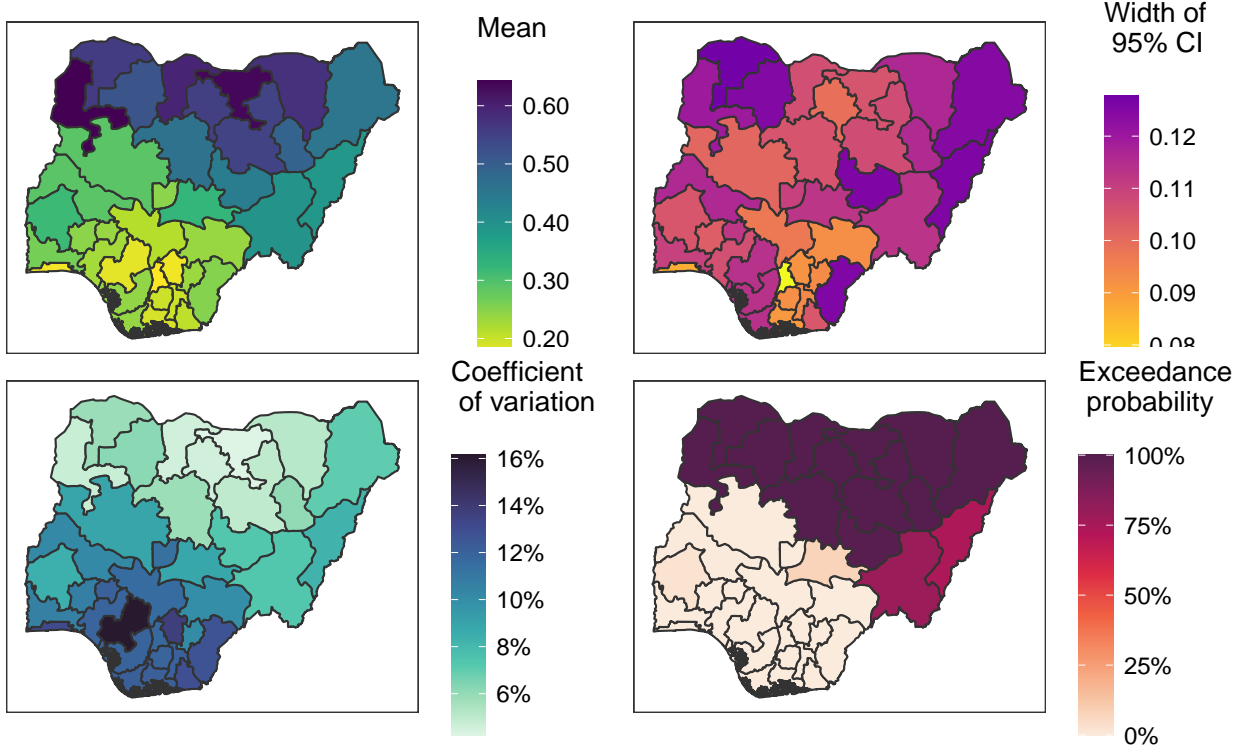

9

# Sample Report from the sae4health R shiny App

Key Statistics Maps: Unit–level model at Admin–2 level

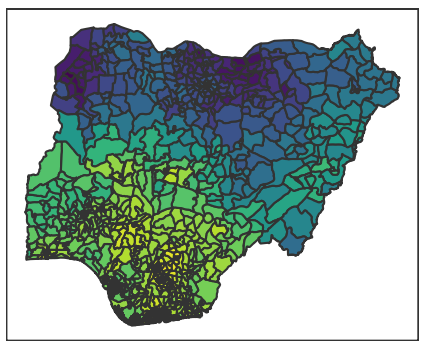
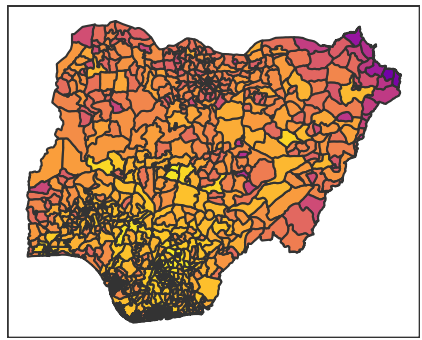

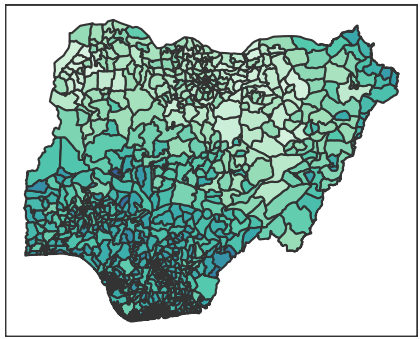
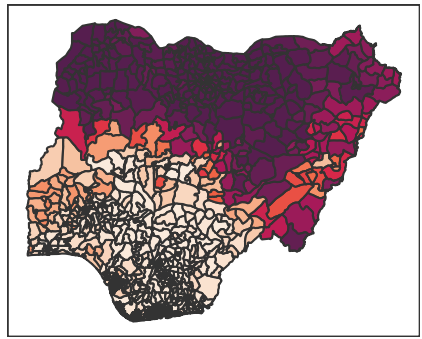

10

# Sample Report from the sae4health R shiny App

Model Comparison: Direct Estimate vs Area−level model at Admin−1 level

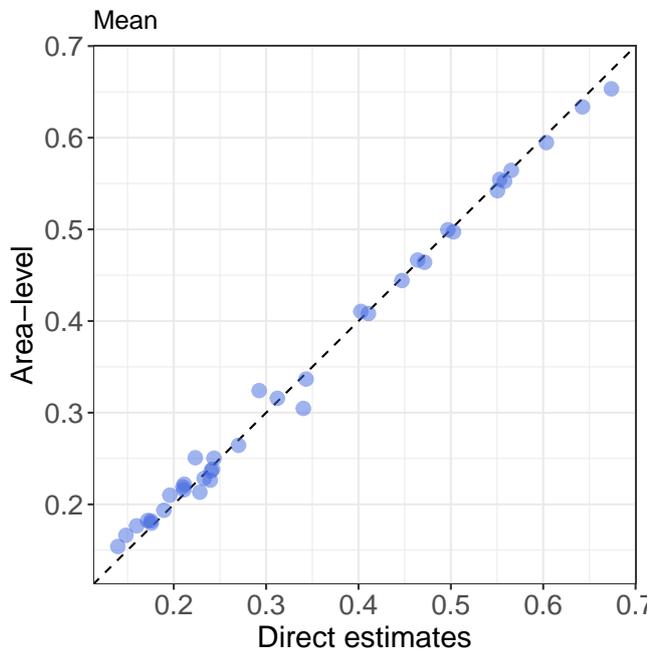

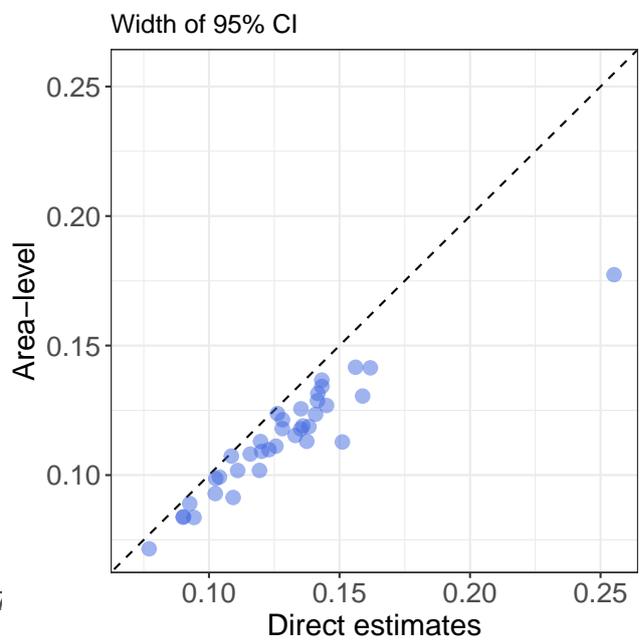

11

# Sample Report from the sae4health R shiny App

Model Comparison: Direct Estimate vs Unit−level model at Admin−1 level

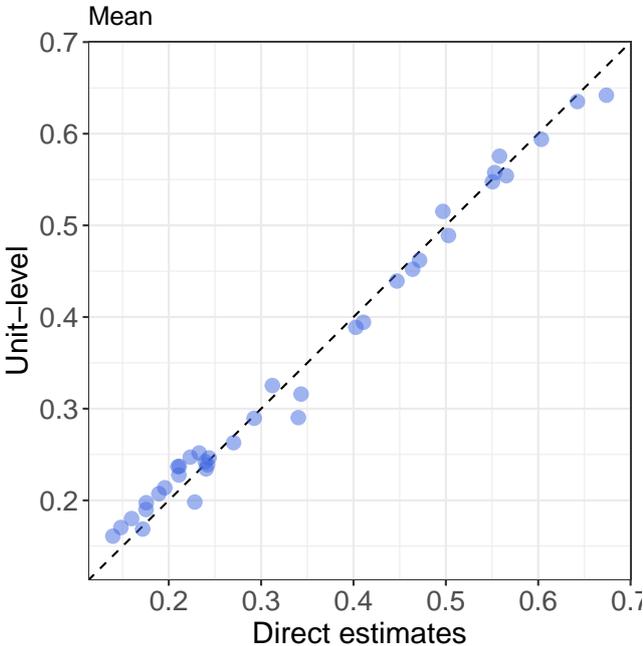
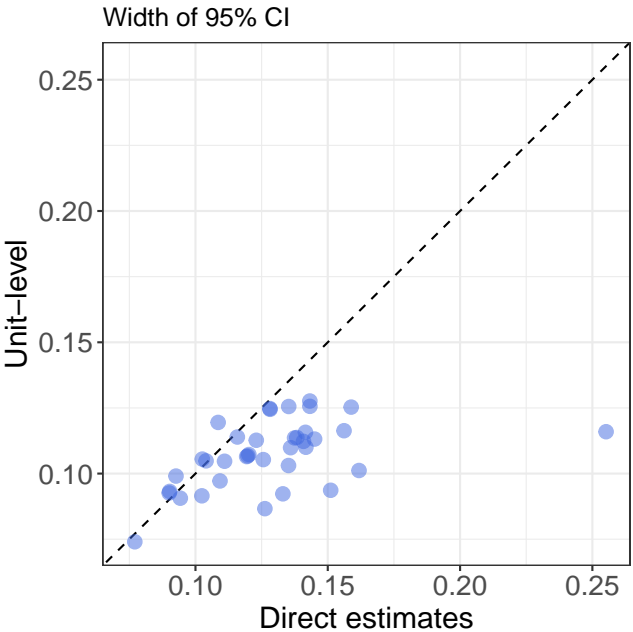

# Sample Report from the sae4health R shiny App

Model Comparison: Direct Estimate vs Unit−level model at Admin−2 level

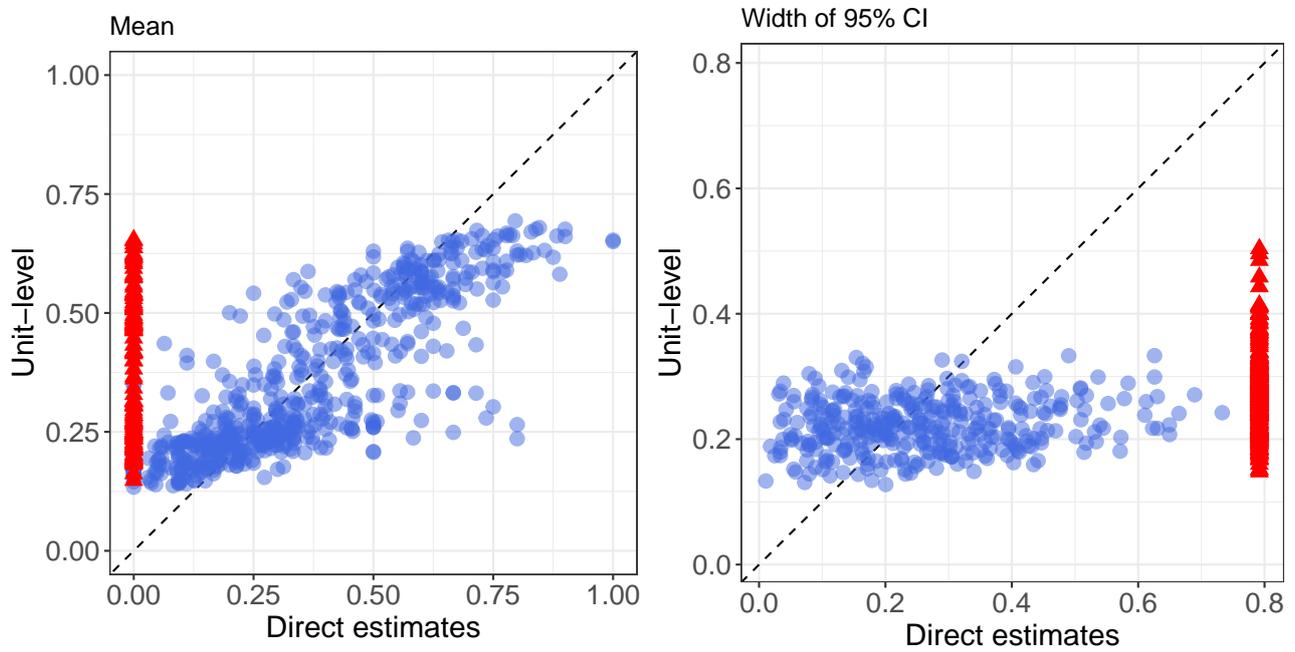



\bibliographystyle{natbib} 
\bibliography{refs}

\end{document}